\documentclass[iop]{emulateapj}

\usepackage{natbib,epsfig,siunitx,url,graphicx,amsmath,footnote,float,xcolor,cases}
\usepackage[colorlinks=true,linkcolor=blue,citecolor=blue]{hyperref}

\begin{document}

\submitted{The Astronomical Journal; accepted}

\title{Orbital precession in the distant solar system; further constraining the Planet Nine hypothesis with numerical simulations}

\author{Matthew S. Clement\altaffilmark{1}, \& Nathan A. Kaib\altaffilmark{2}}

\altaffiltext{1}{Earth and Planets Laboratory, Carnegie Institution
for Science, 5241 Broad Branch Road, NW, Washington, DC
20015, USA}
\altaffiltext{2}{HL Dodge Department of Physics Astronomy, University of Oklahoma, Norman, OK 73019, USA}
\altaffiltext{*}{corresponding author email: mclement@carnegiescience.edu}

\setcounter{footnote}{0}
\begin{abstract}
The longitudes of perihelia and orbital poles of the solar system's dozen or so most remote detected objects are clustered in a manner inconsistent with that of a random sample of uniformly distributed orbits.  While small number statistics and observational biases may explain these features, the statistical significance of the clustering has led to the recent development of the ``Planet Nine hypothesis.''  In the proposed scenario, orbits in the distant solar system are shepherded via secular perturbations from an undetected massive planet on an eccentric orbit.  However, the precession of perihelia and nodes in the outer Kuiper Belt and inner Oort Cloud are also affected by the the giant planets, passing stars, and the galactic tide.  We perform a large suite of numerical simulations designed to study the orbital alignment of Extreme Trans-Neptunian Objects (ETNOs) and Inner Oort Cloud Objects (IOCOs).  In our various integrations that include Planet Nine, we consistently find that $\gtrsim$60$\%$ of ETNOs and IOCOs that are detectable after 4 Gyr are also anti-aligned in perihelia with the distant massive perturber.  However, when we randomly select 17  objects from this sample of remaining orbits, there is significant scatter in the degree of longitude of perihelion and orbital pole clustering that might be observed.  Furthermore, we argue that, in the absence of Planet Nine, 17 randomly drawn orbits should still exhibit some clustering even if the underlying distribution is uniform.  Thus, we find that still more ETNO and IOCO detections are required to confidently infer the presence of Planet Nine.
\end{abstract}

\section{Introduction}

Detached Extreme Trans-Neptunian Objects (ETNOs; 40$\lesssim q \lesssim$ 50 au) and Inner Oort Cloud Objects (IOCOs; $q \gtrsim$ 50 au) are two of the most distant classes of detected bodies in the solar system.  While ETNOs can interact weakly with the four giant planets over Gyr timescales, IOCOs do not \citep{gladman02,bannister27}.  Beyond the realm of IOCOs, Oort cloud bodies with apheila greater than around a few thousand au are more affected by the galactic tide and passing stars than they are by solar system dynamics \citep{kaib09}.  For a recent review of orbital dynamics in the trans-Neptunian regime see \citet{saillenfest20_review}.

As the catalog of these extreme objects grew over the past two decades (particularly notable were the discoveries of IOCOs like Sedna \citep{sedna} and 2012 VP113 \citep{vp113}), it became apparent that their orbital shapes are not distributed uniformly.  This led \citet{vp113} to propose that orbits in the distant solar system might be actively perturbed by a distant, massive planet.  Commonly referred to as Planet Nine, Planet X, or simply the ``perturber;'' this hypothesis \citep{batygin16} spurred an extensive literary output over the last several years \citep[see recent reviews in:][]{batygin19_rev,morb_nes_rev19}.  Work by \citet{bailey16}  and \citet{gomes17} suggests that the unseen planet's large inclination can reproduce the 6$^{\circ}$ solar obliquity and the Sun's slightly misaligned rotational axis via secular interactions.  Additionally, \citet{batygin16b} argued that the two known retrograde TNOs (2008 KV42 \citep{kv42} and 2011 KT19 \citep{kt19}) can not be explained within the conventional framework of the ``Nice Model'' \citep{Tsi05,gomes05,mor05,levison08}, and were likely excited from the scattered disk by an external perturber.  In a similar manner, high inclination Kuiper belt objects (KBOs) that are difficult to generate with conventional evolutionary models \citep{brasser12} might be a consequence of Planet Nine driving the perihelia of scattered objects inward via secular excitation \citep{batygin16b,batygin_morby17,li18}.  Furthermore, multiple studies have investigated the assembly of the Kuiper Belt's orbital structure within the Planet Nine hypothesis \citep[e.g.:][]{nesvorny17,khain18,kaib19}.

\citet{malhotra16} proposed that the four most extreme objects' orbital periods might form a mutual resonant chain with the perturber. However, subsequent work by \citet{bailey18} ruled out the possibility of a resonance based search \citep[e.g.:][]{millholland17} because of the multiplicity of overlapping mean motion resonances (MMRs) and the possibility of chaotic transfer between resonances \citep{becker17}.  Thus, despite extensive study, the physical properties of the proposed massive perturber remain elusive.  

While the parameter space of possible masses and orbits is exhaustive, a sub-Neptune massed planet with moderate inclination seems to best reproduce the observed signature.  \citet{batygin16} performed a detailed analysis of the applicable parameter space \citep[along with subsequent analysis by][]{brown16} and favored a $\sim$10 $M_{\oplus}$ planet with a perihelion of $\sim$150 au, relatively high eccentricity ($e\simeq$ 0.6) and moderate inclination ($i\simeq$30$^{\circ}$).  However, \citet{kaib19} found that the scattered object inclination distribution in simulations including the Oort Cloud and Planet Nine on such an orbit are inconsistent with that of the OSSOS dataset at the $\sim$99.999$\%$ level (regardless of absolute magnitude distribution).  \citet{batygin19_rev} preferred a slightly smaller ($\sim$5 $M_{\oplus}$) perturber on a more circular ($e\simeq$ 0.25, $i\simeq$ 20$^{\circ}$), closer ($a\simeq$ 500 au) orbit.  However, it should be noted that these estimates are somewhat hindered by the fact that the exact underlying distribution of orbital parameters (particularly inclination) in the distant Kuiper Belt and Inner Oort are still unknown.

As the hypothetical perturber remains undetected, the Planet Nine debate continues to center on questions of biases induced by small number statistics \citep{brown16,shankman17} and survey coverage \citep[e.g.:][]{petit11,petit17,alexandersen16,bannister18}.  Unfortunately, given the number of different surveys, the precise degree of bias is difficult to quantify \citep[see, for example,][for a detailed discussion]{kavelaars19}.  In this manuscript, we present a large suite of simulations of ETNO and IOCO stability in the presence of the hypothetical Planet Nine that aim to infer the degree of expected orbital clustering.  Our numerical integrations include algorithms \citep{kaib_quinn_sci_09,kaib19} designed to account for the effects of stellar encounters \citep{fernandez00,rickman08} and the galactic tide \citep{heisler86,dones04}.  We follow a de-biasing approach similar to that presented in \citet{brown17} and \citet{brown19}.  We first assume that the underlying distribution of longitudes of perihelia ($\varpi$) and orbital poles (longitude of ascending node: $\Omega$) of all KBOs in the Minor Planet Center (MPC) database is representative of the cumulative longitudinal survey bias.  We then scrutinize whether bias-informed simulated detections drawn from our evolved populations of extreme objects exhibit orbital clustering consistent with what is observed.  Since large swaths of orbital parameter space in the distant solar system have been shown to be chaotic over Gyr-timescales \citep{saillenfest19}, we also investigate the stability of clones of the 8 most extreme objects individually.  Finally, we conclude our study by briefly addressing the possibility of the observed clustering in longitudes being the result of a ``chance alignment.''  Specifically, we aim to characterize the diffusion timescale for a coincidentally aligned population, and comment on the likelihood of inferring a significance in clustering with alignment longitude as a free parameter.

 \section{Methods}
 
 \subsection{Numerical simulations}
 For our numerical simulations, we use a version of the $SWIFT$ N-body package \citep{levison94} that is modified to include the effects of the galactic tide and stellar encounters \citep{kaib19}.  The galactic tide is modeled with a radial component based on the Oort Constants \citep{oort27}, and a vertical term derived from the local density of matter in the Milky Way's disk \citep[fixed at 0.1 $M_{\odot} \medspace pc^{-3}$ in our simulations:][]{levison01}.  Stellar encounters are generated at random assuming the Present Day Mass Function of \citet{reid02_PDMF}, local velocity dispersions of \citet{rickman08}, and a stellar density of 0.034 $M_{\odot} \medspace pc^{-3}$ \citep{kaib19}.  Each stellar encounter is integrated directly for $r<$1 pc.  While these parameters have not remained constant in the solar neighborhood over the past 4 Gyr \citep[e.g.:][]{brasser06,kaib11}, we select this simplified model for a first order approximation of the dynamic environment in the distant solar system.

\begin{table*}
\centering
\begin{tabular}{c c c c c c}
\hline
Name & Source & Mass ($M_{\oplus}$) & $a$ (au) & $e$ & $i$ ($^{\circ}$) \\
\hline
P9a & \citet{batygin16} & 10.0 & 700 & 0.6 & 30.0 \\
P9b & \citet{batygin19_rev} & 5.0 & 500 & 0.25 & 20.0 \\
\hline
\end{tabular}
\caption{Initial orbits for Planet Nine used in our simulations.  Angular orbital elements are selected at random.}
\label{table:p9}
\end{table*}
 
 Each of our dynamical simulations include the four giant planets on their modern orbits, the hypothetical Planet Nine, and 1,000 massless test particles.  For completeness, we test two different proposed orbits \citep{batygin16,batygin19_rev} for Planet Nine (table \ref{table:p9}).  Angular orbital elements for our test particles (specifically argument of perihelon, longitude of ascending node, and mean anomaly; $M$) are drawn randomly from uniform distributions, while inclinations are selected from the following distribution \citep[in order to resemble the modern distribution of KBOs, e.g.:][]{brown01,kaib16}:
 
 \begin{equation}
 	f(i) = sin(i)exp\Big(-\frac{i}{2\sigma_{i}^{2}}\Big)
 	\label{eqn:fi}
 \end{equation}
 
 Our simulations are designed to study the evolution of orbital clustering of ETNOs, IOCOs (table \ref{table:ics}) and clones (table \ref{table:ics2}) of specific significant \citep[e.g.:][]{sheppard19} TNOs under the influence of the external perturber \citep{batygin16}.  For our simulations aiming to investigate the bulk populations of ETNOs and IOCOs, we select semi-major axes and perihelia randomly from the ranges given in table \ref{table:ics}.  We also perform one set of control simulations that do not include Planet Nine, and make use of the same initial conditions as in our ETNO,low-$i$ batch.  For our object clone runs, each test particle is randomly assigned a value within 1 au of the object's actual semi-major axis, 0.5 au of its pericenter, and 0.5$^{\circ}$ of its inclination (table \ref{table:ics2}).  We then perform ten, 4 Gyr simulations of each initial condition set.
 
 \begin{table}
 	\centering
 	\textbf{Bulk ETNO/IOCO simulations}
 	
 	\begin{tabular}{c c c c}
 	\hline
	Set & $a$ (au) & $q$ (au) & $\sigma_{i}$ ($^{\circ}$) \\
	\hline
	Control & 300-500 & 40-50 & 1.0 \\
 	ETNO,low-$i$ & 300-500 & 40-50 & 1.0 \\
 	ETNO,mod-$i$ & 300-500 & 40-50 & 10.0 \\
 	ETNO,high-$i$ & 300-500 & 40-50 & 20.0 \\
 	IOCO,low-$i$ & 800-1200 & 50-100 & 1.0 \\
 	IOCO,mod-$i$ & 800-1200 & 50-100 & 10.0 \\
 	IOCCO,high-$i$ & 800-1200 & 50-100 & 20.0 \\
 	\hline
 	\end{tabular}
	\caption{Summary of semi-major axes, perihelia and $\sigma_{i}$ (equation \ref{eqn:fi}) for TNO test particles in our respective simulation sets.}
	\label{table:ics}
 \end{table}

 \begin{table}
 	\centering
 	\textbf{Object Clone simulations}

 	\begin{tabular}{c c c c c}
	\hline
 	Class & Object & $a$ (au) & $q$ (au) & $i$ ($^{\circ}$) \\
 	\hline
 	IOCOs &2012 VP113 & 266 & 80.3 & 24.1 \\
 	& Sedna & 507 & 76.0 & 11.9 \\
 	& 2015 TG387 & 1200 & 65 & 11.7 \\
 	\hline
 	ETNOs & 2013 SY99 &730 & 49.9 & 4.2 \\
 	& 2010 GB174 & 351 & 48.8 & 21.5 \\
 	& 2014 SR349 & 299 & 47.6 & 18.0 \\
 	& 2004 VN112 & 327 & 47.3 & 25.6 \\
 	& 2015 RX245 & 430 & 45.5 & 12.2 \\
 	\hline
 	\end{tabular}
 	\caption{Summary of barycentric semi-major axes, perihelia and inclinations (taken from the $MPC$ online database) for 8 detached TNOs with the largest perihelia.}
	\label{table:ics2}
 \end{table}
 
 \subsection{De-biasing prescription}
 \label{sect:bias}
 
\begin{figure}
 	\centering
 	\includegraphics[width=0.5\textwidth]{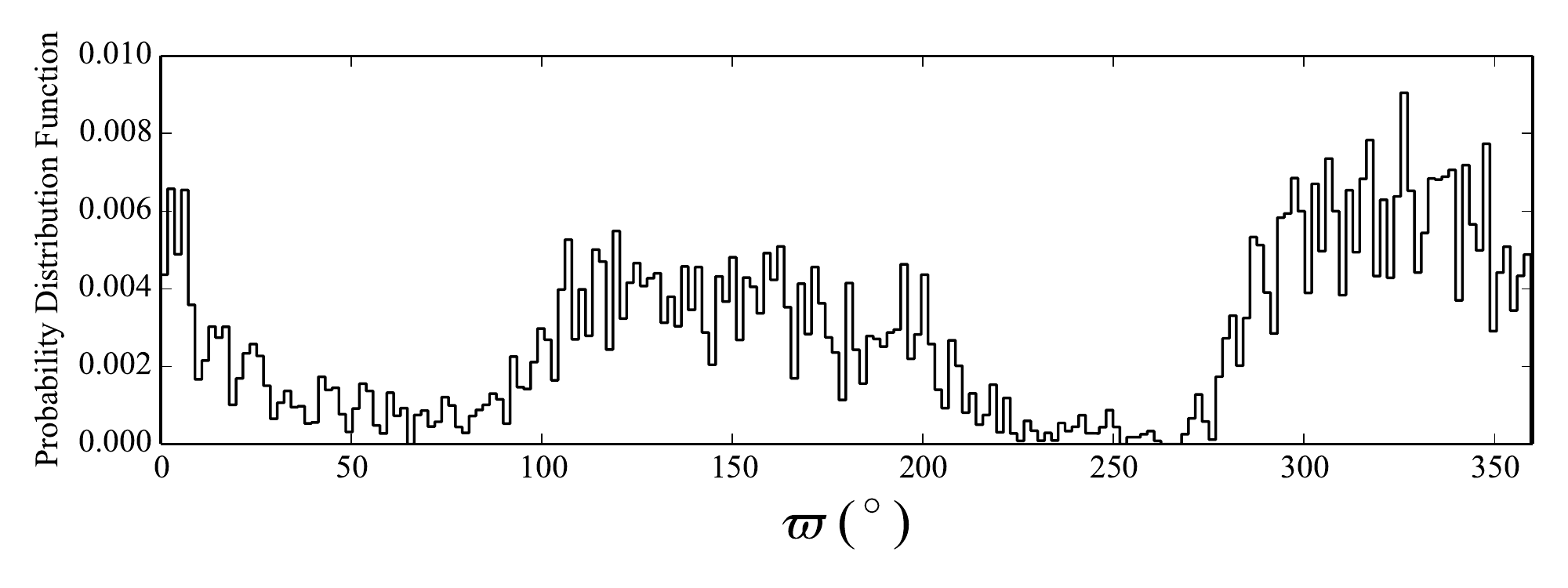}
 	\caption{Example $\varpi$ bias function for Sedna \citep[e.g.: figure 1 in][]{brown17}}
 	\label{fig:sedna}
 \end{figure}

 We follow a de-biasing scheme similar to that of \citet{brown17} and \citet{brown19} for determining the cumulative bias with respect to $\varpi$ and $\Omega$ for survey detections in the distant solar system.  This method makes the fundamental assumption that the probability of detecting a TNO at a given $\varpi$ or $\Omega$ can be inferred from the complete $MPC$ catalog of KBOs with $q\geq$ 30 au (for this work queried on 18 December 2019).  We first calculate the ecliptic latitude, longitude, and brightness for each KBO at the time it was discovered.  This can be considered a sample of survey times, limiting magnitudes and positions on the sky that might have been capable of detecting one of our distant TNOs, had it possessed a different set of orbital angles.  To calculate the longitude of perihelion bias for a given object of interest (see section \ref{sect:objects}), we iterate through each possible $\varpi$/$M$ combination (assuming a uniform distribution of angles) for each KBO detection in our aforementioned ``sample of surveys.''  Through this process we determine the range of $\varpi$ and $M$ for which each extreme object would be bright enough, and at the correct position (within 1$^{\circ}$), to be ''detected'' by any of the surveys.  This allows us to build a probability distribution of detection with respect to $\varpi$ for each distant TNO (an example for Sedna is plotted in figure \ref{fig:sedna}).  As discussed in more depth in \citet{brown17}, we also attempt to account for several  of our fundamental assumption's fallacies \citep[though it should be noted that][determined that the final de-biased distributions do not depend strongly on these assumptions]{brown19}.  Specifically, these are related to the over-abundance of resonant plutinos detected near pericenter, the dependency of detection latitude on the KBO inclination distribution, and the fact that all detection surveys are not sensitive to extremely faint objects.  In practice, we account for these inherent biases by limiting the underlying population of KBOs to those with $a>$ 40 au (thus removing the plutinos), weighting each detection by the expected density of KBOs at that latitude \citep[by converting equation \ref{eqn:fi} to a latitudinal distribution, assuming $\sigma_{i}=$14.9:][]{brown01} and only considering objects ``detectable'' at heliocentric distances $<$ 90 au.  Our method for determining the bias in $\Omega$ follows in a similar manner.  In that case, simulated detections have fixed values of $\varpi$, and $\Omega$ is the angle that is iterated over.  Our results should be interpreted with an understanding that our de-biasing scheme is inherently flawed, but still a reasonable approach to capture the cumulative bias that is entangled in decades of different surveys utilizing various methodologies and reporting differing parameters.
 
 \subsection{Objects selected}
 \label{sect:objects}
 
 At this point it is worth discussing the selection criteria used by authors analyzing orbital clustering in the distant solar system.  Recent work by \citet{brown19} and \citet{batygin19_rev} use delimiting values of $q>$ 30 au, $a>$ 250 au, and $i<$ 40$^{\circ}$; 14 objects.  However, this excludes the high inclination ($i=$ 54$^{\circ}$, $q=$ 35 au, $a=$ 863 au) object 2015 BP519 discovered by the Dark Energy Survey \citep{darkenergy16}, the extreme orbit of which seems to be explained by interactions with Planet Nine \citep{becker18}.  Additionally, the object uo5m93 reported in the full OSSOS data release \citep[Outer Solar System Origins Survey:][]{bannister18} would also meet these criteria, though the number of observations was insufficient to adequately constrain its orbit in accordance with the standards established by OSSOS. Finally, two additional objects' orbits have been determined and added to the $MPC$ database since the work of \citet{brown19} and \citet{batygin19_rev}: 2013 RA109 ($q=$ 46 au, $a=$ 315 au, $i=$ 6.5$^{\circ}$) and 2013 SL102 ($q=$ 35 au, $a=$ 863 au, $i=$ 54$^{\circ}$).  While each of these new objects' longitude of perihelion loosely opposes the proposed value for Planet Nine, they are both at angles near the extreme of the apparent observed clustering
 
 \citet{sheppard19} studies 13 objects that meet a more conservative minimum perihelia value ($q=$ 40 au), and more relaxed semi-major axis limit ($a>$ 150 au).  As a result, that manuscript considers 2013 UT15 ($q=$ 44 au, $a=$ 196 au), 2013 GP136 ($q=$ 41 au, $a=$ 155 au) and 2000 CR105 ($q=$ 44 au, $a=$ 227 au), but does not include the objects with $q\simeq$ 35-38 au (2014 FE72, 2007 TG422, 2013 RF98 and 2015 GT50) addressed by \citet{batygin19_rev}.  While these imposed limits are indeed somewhat arbitrary, it is also readily apparent that bodies with perihelia near $\sim$30 au can interact strongly with Neptune.  The degree to which Neptune perturbs these low-$q$ objects decreases with increasing semi-major axis \citep{morby08}.  Thus, as an objects' dynamical separation from Neptune is of most importance when considering orbital shepherding by an external perturber, we focus our dynamical simulations on detached objects satisfying the more conservative minimum perihelion distance of 40 au (table \ref{table:ics}).  We take this one step further in our simulations designed to study object clones (table \ref{table:ics2}), and select only objects with $q>$ 45 au (thus the objects in common to both the listings of \citet{sheppard19} and \citet{batygin19_rev}).  For consistency, when we compare our simulation-generated systems of clustered orbits to the observed objects using the test of \citet{brown19}, we employ a similar object selection criteria: $q>$ 30 au and $a>$ 250 au.  Therefore, with the inclusion of 2015 BP519, 2013 RA109 and 2013 SL102, our study considers the observed orbital clustering of 17 objects.

\section{Results and Discussion}

\subsection{Distinguishing between a Planet Nine distribution and a uniform distribution}
\label{sect:disc}

\begin{figure*}
	\centering
	\includegraphics[width=.4\textwidth]{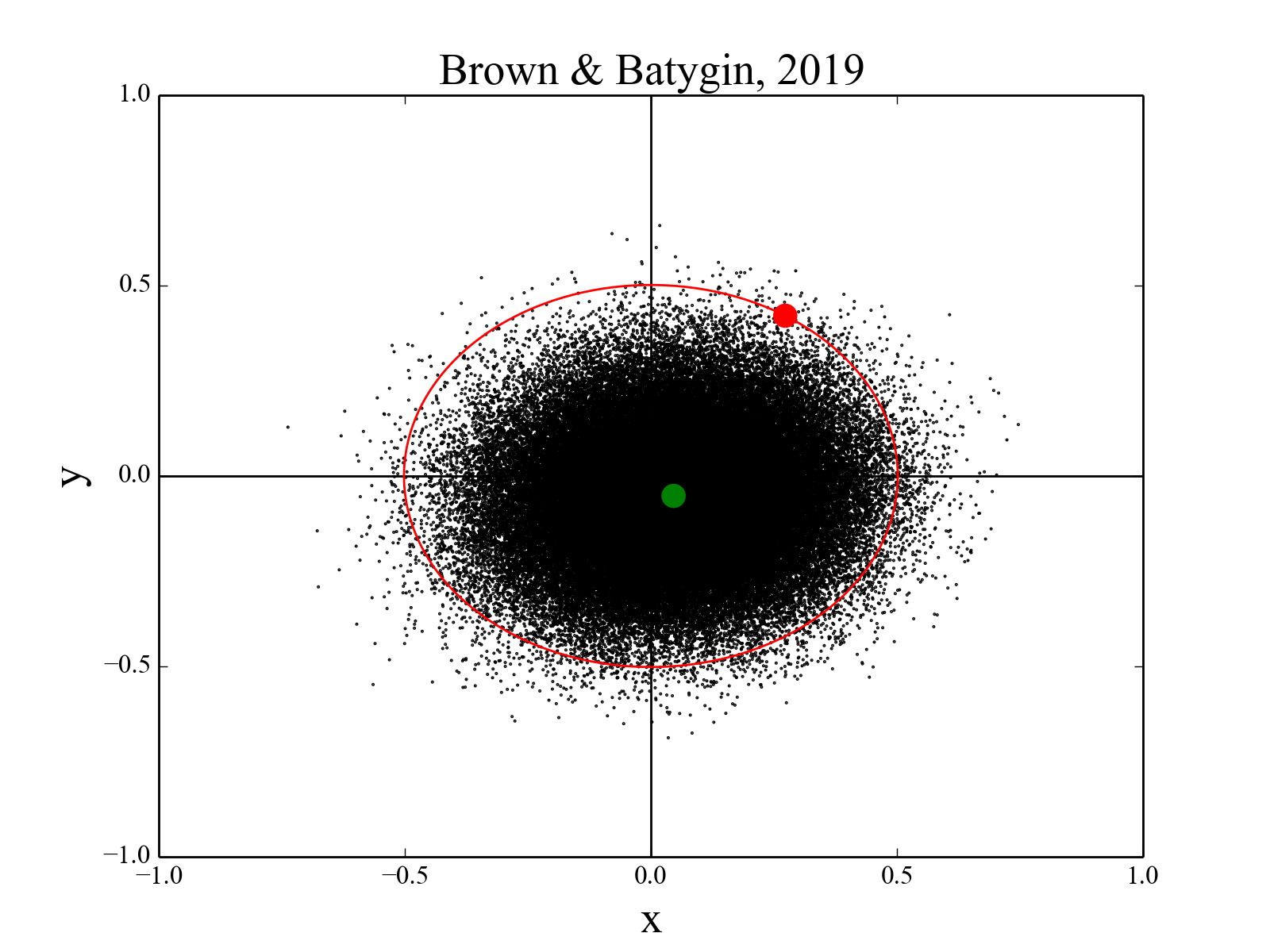}
	\includegraphics[width=.4\textwidth]{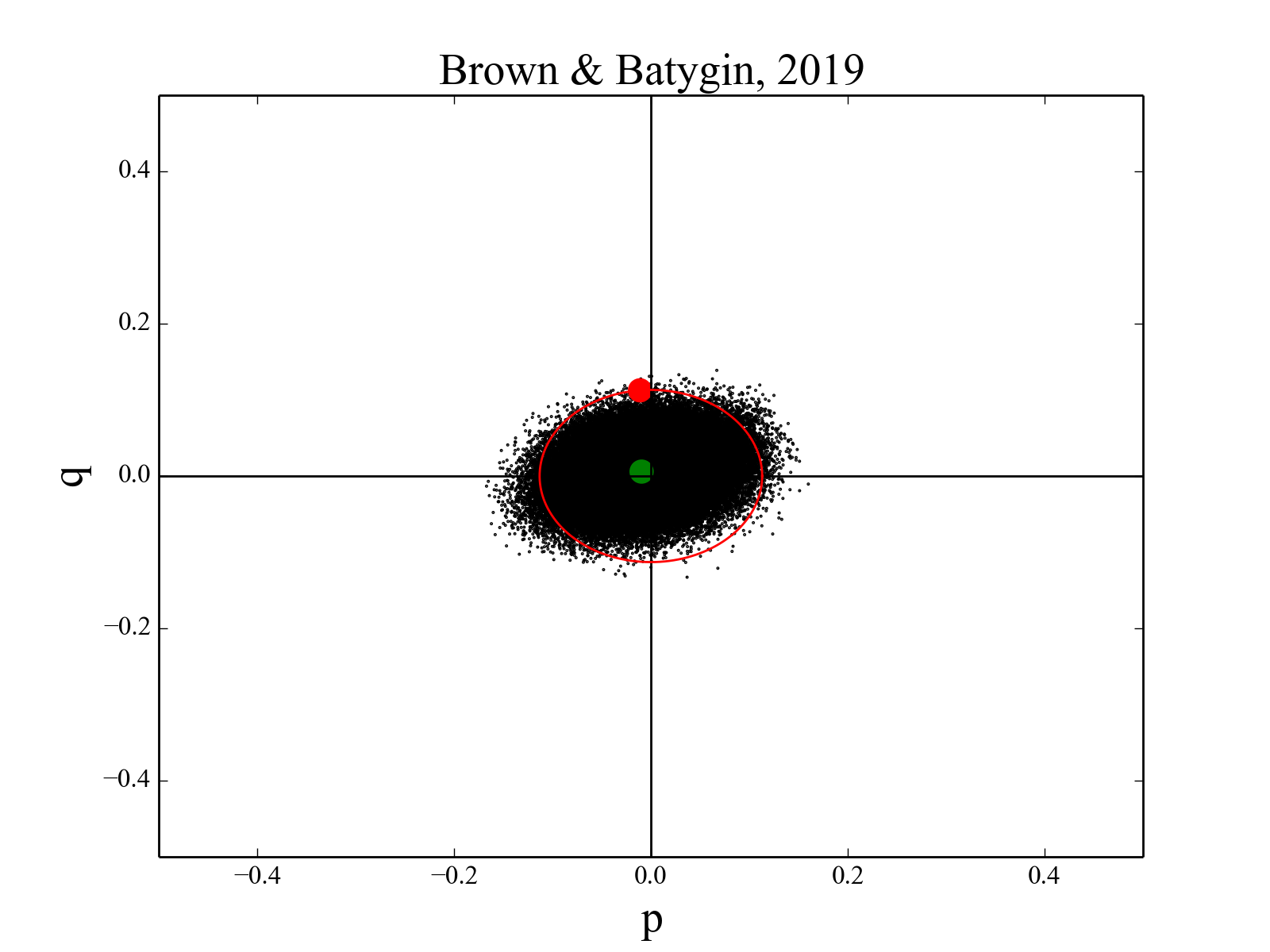}
	\qquad
	\includegraphics[width=.4\textwidth]{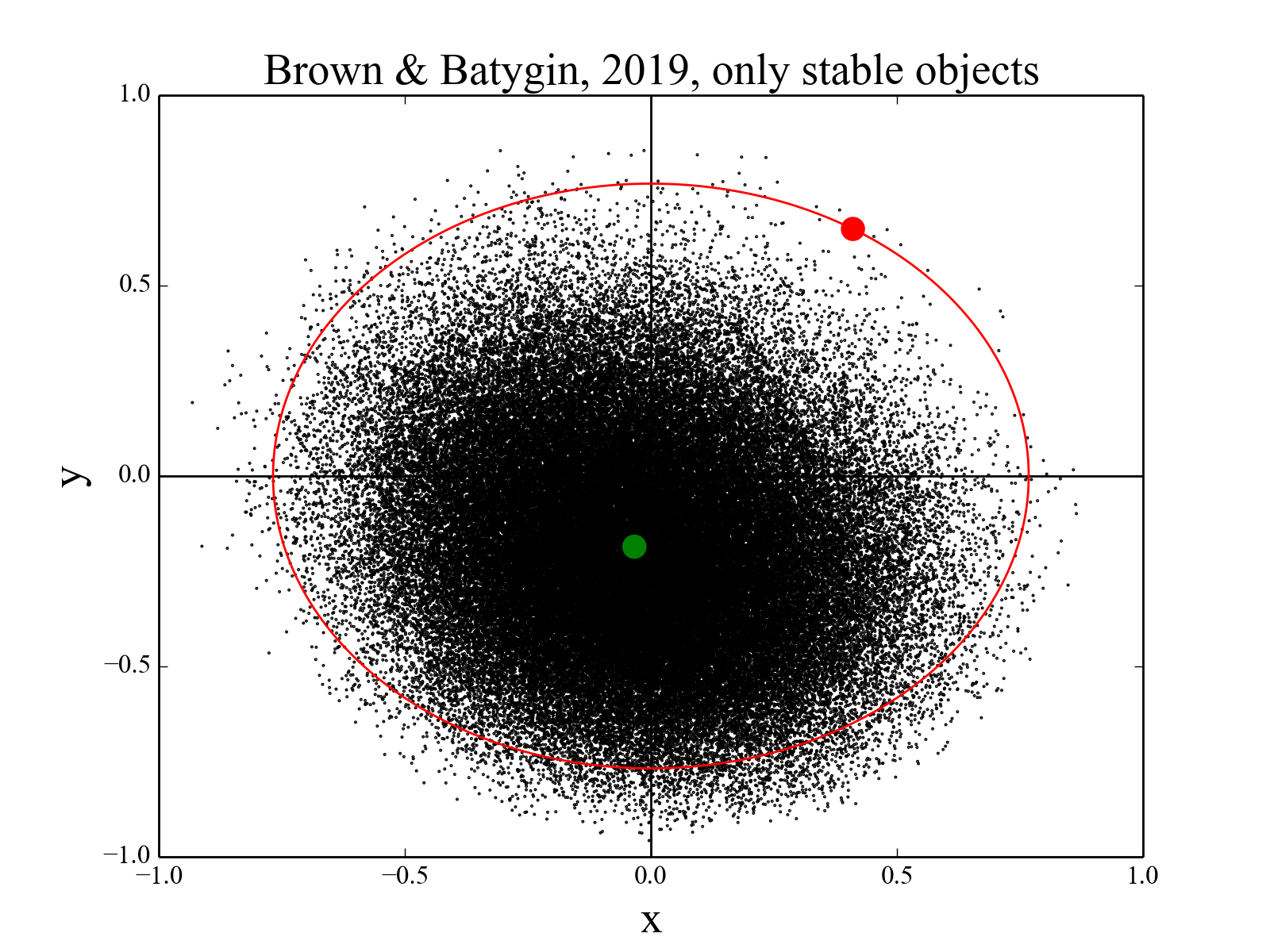}
	\includegraphics[width=.4\textwidth]{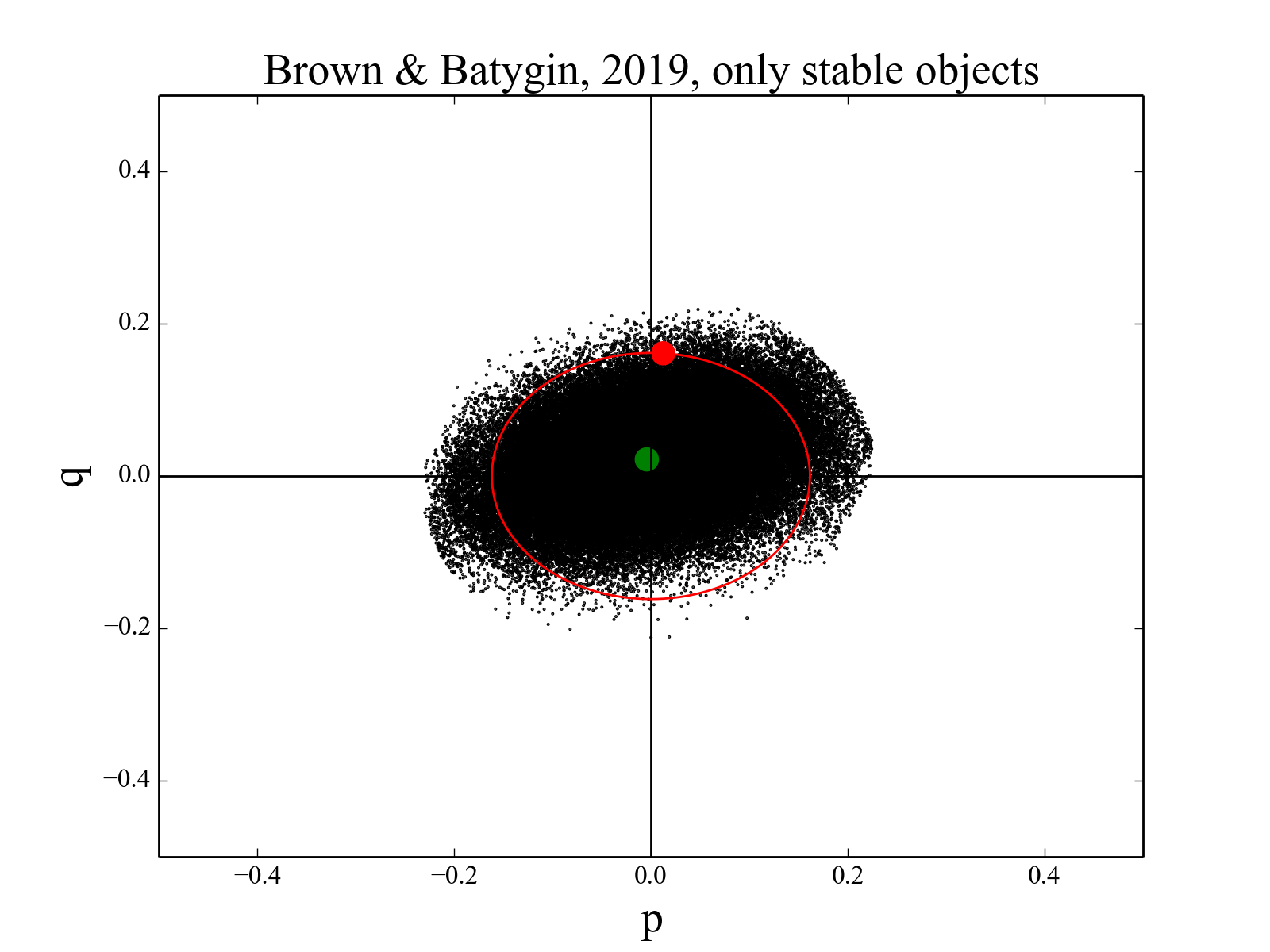}	
	\qquad
	\includegraphics[width=.4\textwidth]{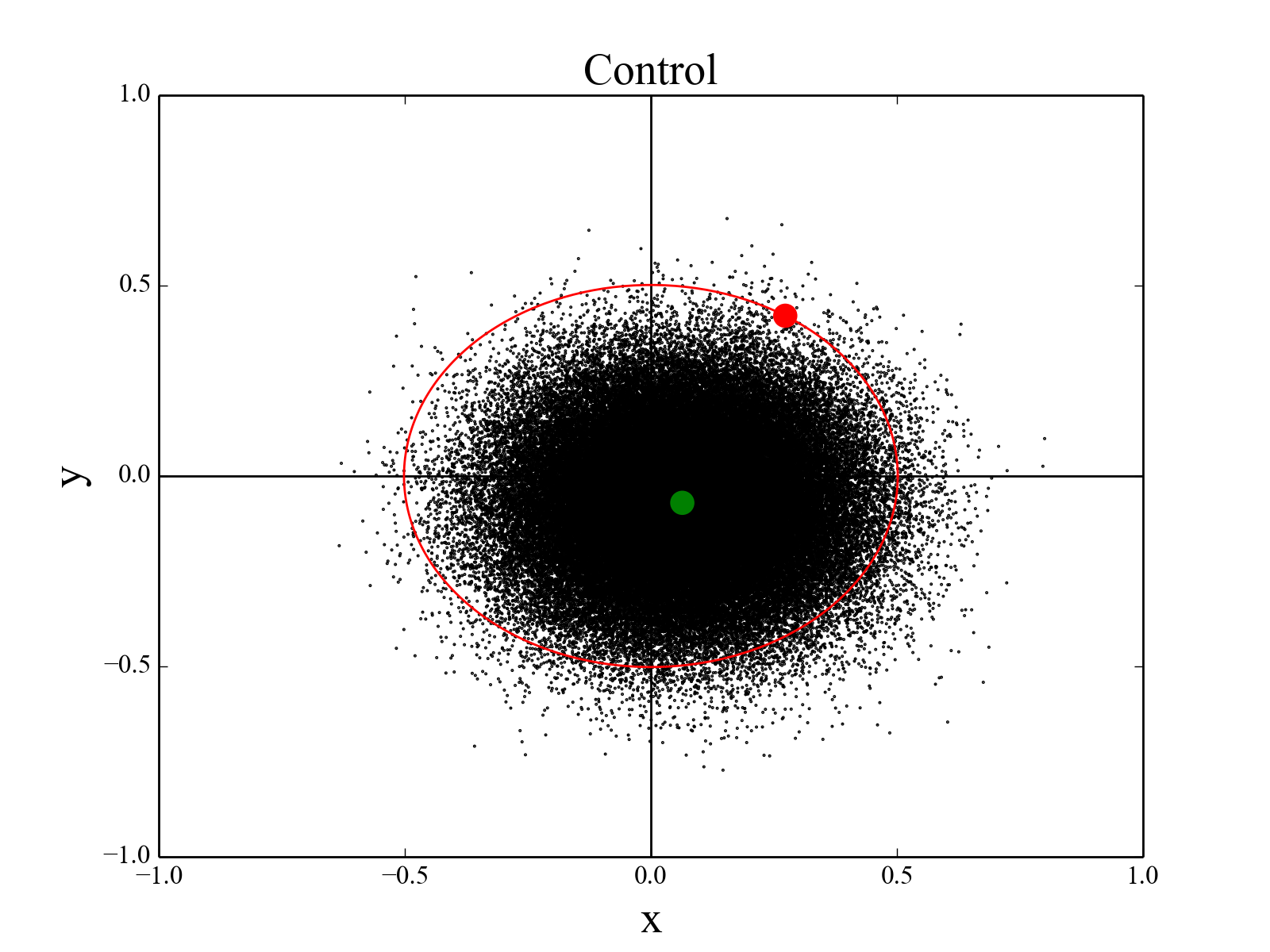}
	\includegraphics[width=.4\textwidth]{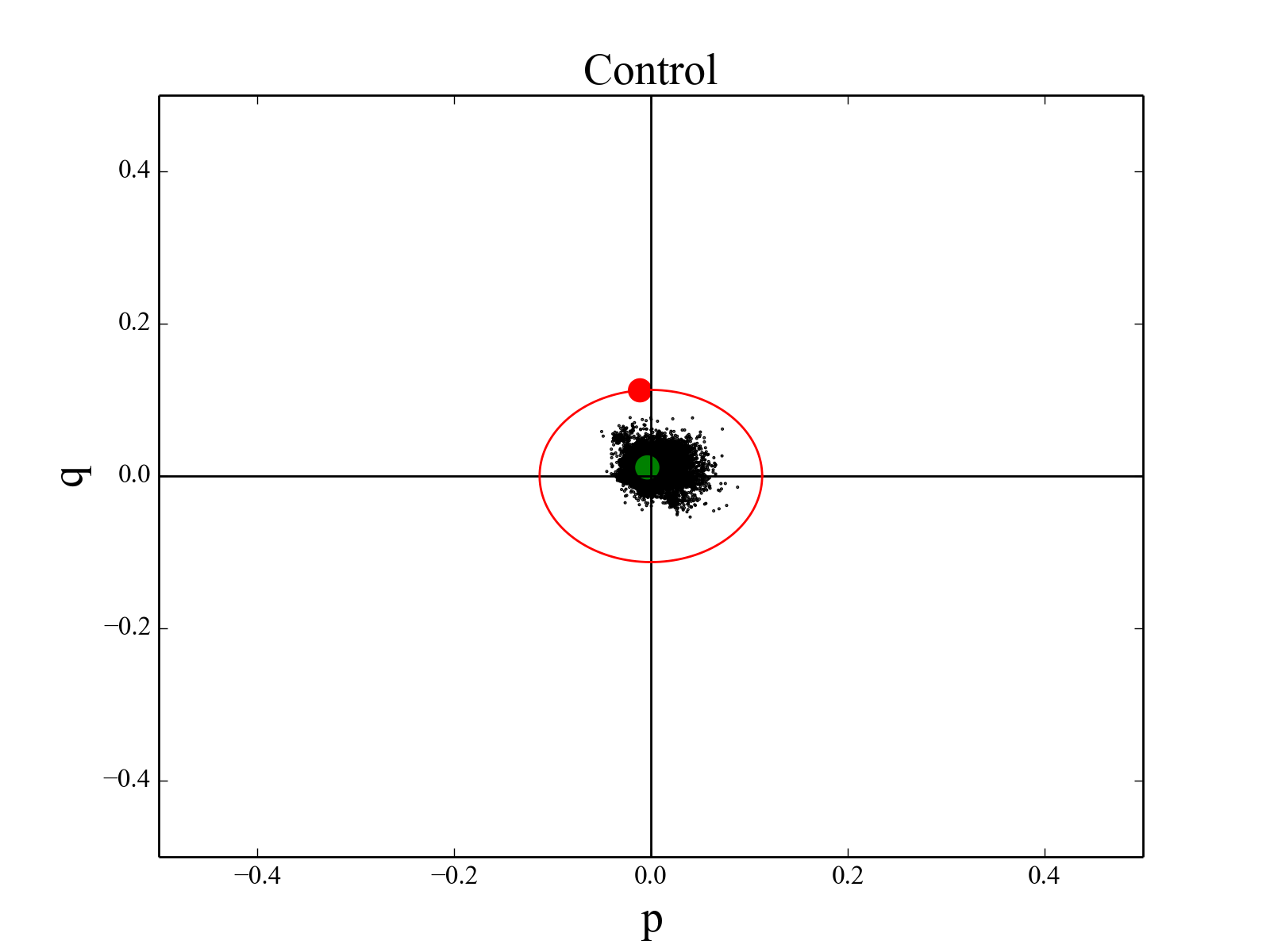}	
	\caption{Top Panels: Reproduction of figures 1 and 2 from \citet{brown19} displaying the observed clustering of 17 orbits in the distant solar system (red point) in $x/y$ and $p/q$ space, compared with 100,000 iterations (black points, average plotted in green) where each object's longitudes of perihelia is drawn randomly from our generated bias distribution functions.  1.4$\%$ and 2.2$\%$ of the random iterations are more strongly clustered than what is observed (red circles) in $x/y$ and $p/q$ space, respectively.  Middle Panels: The same as the above except only considering the 6 objects classified by \citet{batygin19_rev} as having dynamical lifetimes greater than the age of the solar system.  2.4$\%$ and 9.2$\%$ of the random iterations are more strongly clustered than what is observed (red circles) in $x/y$ and $p/q$ space, respectively.  Bottom Panels: The same plot as in the top panels except, here, we randomly simulate 17 detections from each of our Control, simulations that do not include Planet Nine (calculating detection bias in the same manner).}
	\label{fig:brown19}
\end{figure*}

With few exceptions, consistent with \citet{batygin19_rev}, we find our P9b  configurations (5.0 $M_{\oplus}$ planet with $a=$ 500 au, $e=$ 0.25, and $i=$ 20.0$^{\circ}$) to be significantly more successful than the corresponding P9a  sets (10.0 $M_{\oplus}$ planet with $a=$ 700 au, $e=$ 0.60, and $i=$ 30.0$^{\circ}$).  Thus, we focus the majority of our discussion in the subsequent sections on P9b simulations.  We begin our analysis by repeating the Monte Carlo orbital clustering significance test of \citet{brown19}.  We first calculate the bias distribution function independently for each TNO as described in section \ref{sect:bias}, and then perform a transformation to the orthogonal basis of canonical Poincar\'{e} variables \citep[e.g.:][]{morby02_book}.  Beginning from the modified Delaunay variables:
\begin{flalign}
& \Lambda_{j} = \mu_{j} \sqrt{G(m_{0}+m_{j})a_{j}} \; , & & \lambda_{j} = \omega_{j}+\Omega_{j}+M_{j} \\
& \Gamma_{j} = \Lambda_{j}\big(1-\sqrt{1-e_{j}^{2}}\big) \; , & & \gamma_{j} = -\omega_{j}-\Omega_{j} \\
& Z_{j} =  \Lambda_{j} \sqrt{1-e_{j}^{2}}\big(1-\cos{i_{j}}\big) \; , & & z_{j} = -\Omega_{j}
\end{flalign}
The Poincar\'{e} variables are defined as follows:
	\begin{flalign}
		& x_{j} = \sqrt{2\Gamma_{j}}\cos{-\gamma_{j}} \; , & & y_{j} = \sqrt{2\Gamma_{j}}\sin{-\gamma_{j}} \\
		& p_{j} = \sqrt{2Z_{j}}\cos{-z_{j}} \; , & & q_{j} = \sqrt{2Z_{j}}\sin{-z_{j}} \\
	\end{flalign}
When scaled by the quantity of $\Lambda$, a vector in the $x/y$ plane points in the direction of the longitude of perihelion with magnitude proportional to the eccentricity.  Similarly, vectors in $p/q$ space point towards the orbital node, scale directly with inclination and indirectly, with eccentricity.  The top left panel of figure \ref{fig:brown19} reproduces the test of \citet{brown19} in $x/y$ space.  By performing 100,000 iterations of randomly selecting a longitude of perihelion for each of the 17 distant (this includes three additional objects not considered in that work; see section \ref{sect:objects}), clustered objects from each body's respective bias distribution, we confirm the result of \citet{brown19}.  In $x/y$ space, the observed clustering (red point and circle) falls outside of $\sim$98.3$\%$ of the randomly selected sets of orbits.  In $p/q$ space, the result is similar ($\sim$97.7$\%$).  It is important to point out that the significance of the observed clustering is higher with the addition of 2015 BP519, 2013 RA109 and 2013 SL102.  When we remove these objects, and thus consider the same 14 orbits as \citet{brown19}, we replicate the results of that work ($\sim$96.0$\%$ or iterations outside the red circle in $x/y$ space, and $\sim$96.5$\%$ in $p/q$ space).  In this manner, \citet{brown19} concluded that (by combining the $x/y$ and $p/q$ clustering), the probability that the 14 observed orbits are drawn from a uniform distribution of orientations is just $\sim$0.2$\%$.  However, it should be noted here that the sample of OSSOS \citep{bannister16_ossos} detected objects have been shown to be consistent with originating from a uniform distribution in independent studies \citep[for example:][]{kavelaars19,brown19}.

These results should be taken in the appropriate context given that, of the 17 objects discussed in the previous paragraph, only 6 have dynamical lifetimes significantly longer than the age of the solar system \citep{batygin19_rev}.  Thus, it is likely that several of the 17 detached objects are ``interlopers'' on orbits that coincidentally cluster with those of the other TNOs.  The middle panel of figure \ref{fig:brown19} plots the same test of \citet{brown19} for these 6 objects.  In this case, the observed clustering (red point and circle) falls outside of $\sim$97.6$\%$ and $\sim$90.8$\%$ of the randomly selected sets of orbits in $x/y$ and $p/q$ space, respectively.  While these 6 objects are all highly clustered in $\varpi$ \citep[see, for example, figure 6 of][]{batygin19_rev}, the overall statistical significance is lessened because of the small number of objects.

Next, we ask the same question in reverse.  Given the cumulative biases, to what degree would the orbits of 17 objects detected from our underlying distribution of simulated test particles be clustered in $p/q$ and $x/y$ space?  This is essentially the test used by authors investigating the most probable possible orbit for Planet Nine \citep[e.g.:][]{batygin19_rev}.  As our simulations include the effects of the galactic tide and stellar encounters, we argue that this analysis and comparison is an important step in assessing the strength of the Planet Nine hypothesis, and the success of the most probable orbits.  Furthermore, analyzing the differences between the expected clustering of the different object classes (ETNOs and IOCOs) might provide inferences into the significance of new detections.  

The bottom panel of figure \ref{fig:brown19} depicts this test for our Control simulations.  We find that our Control simulations provide a good match to the expected range of clustering that might be observed in $x/y$ space assuming the sample of detected objects possesses orbital angles drawn from a uniform distribution (top panel).  Thus, perturbations from the galactic tide and stellar encounters do not appear to induce a significant bias in our simulations.  Moreover, this result implies that any bias observed in our simulations that include Planet Nine is a result of perturbations from the hypothetical distant planet.  On the other hand, the tight clustering of our Control orbits in $p/q$ space (bottom right panel of figure \ref{fig:brown19}) is a result of the region's unconstrained inclination distribution's effect on orbital pole clustering.  Since the magnitude of vectors in $p/q$ space scale with inclination, we expect objects from our low-$i$ Control runs to possess $p/q$ values near the origin.

\begin{figure*}
	\centering
	\includegraphics[width=0.4\textwidth]{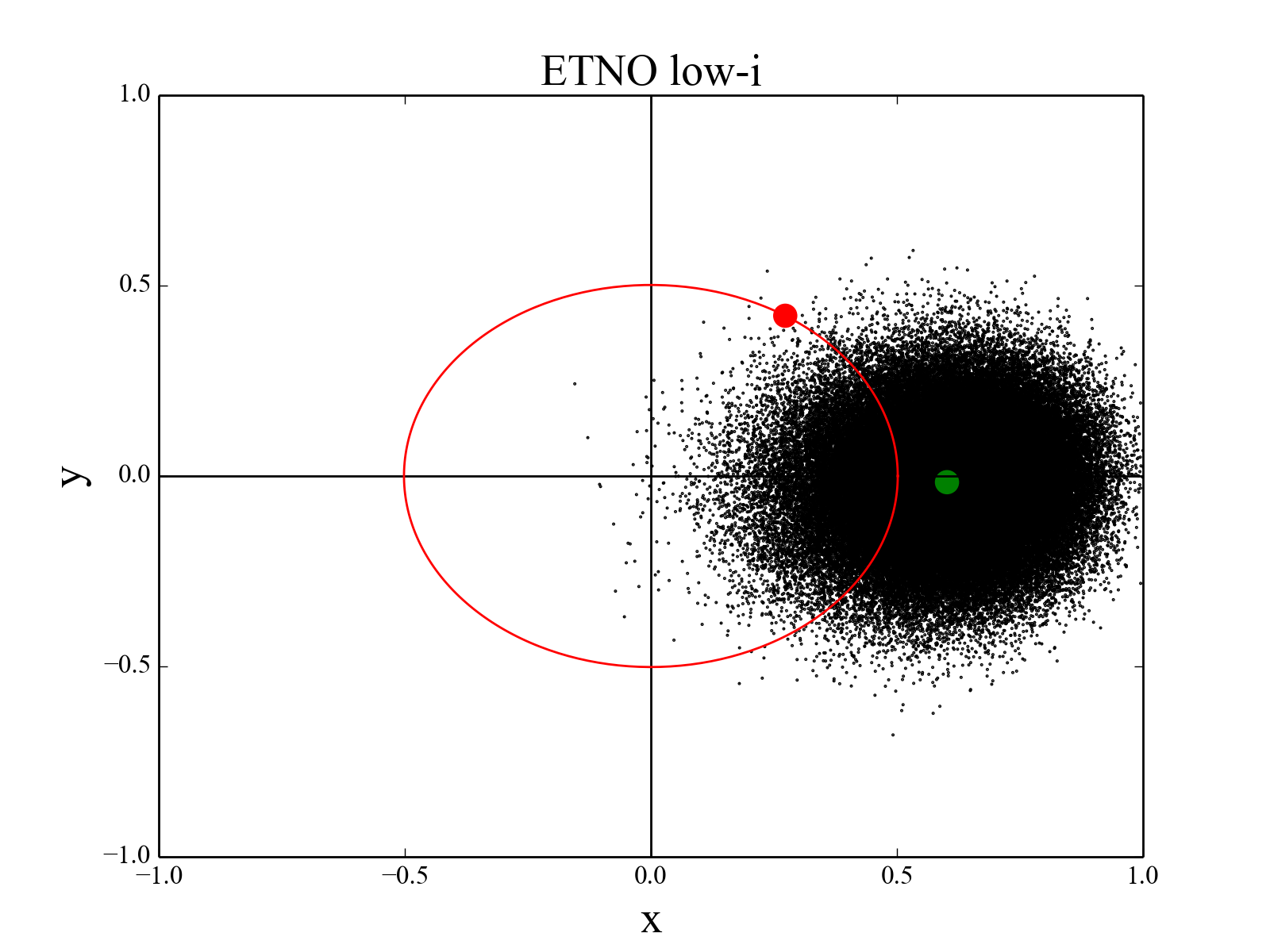}
	\includegraphics[width=0.4\textwidth]{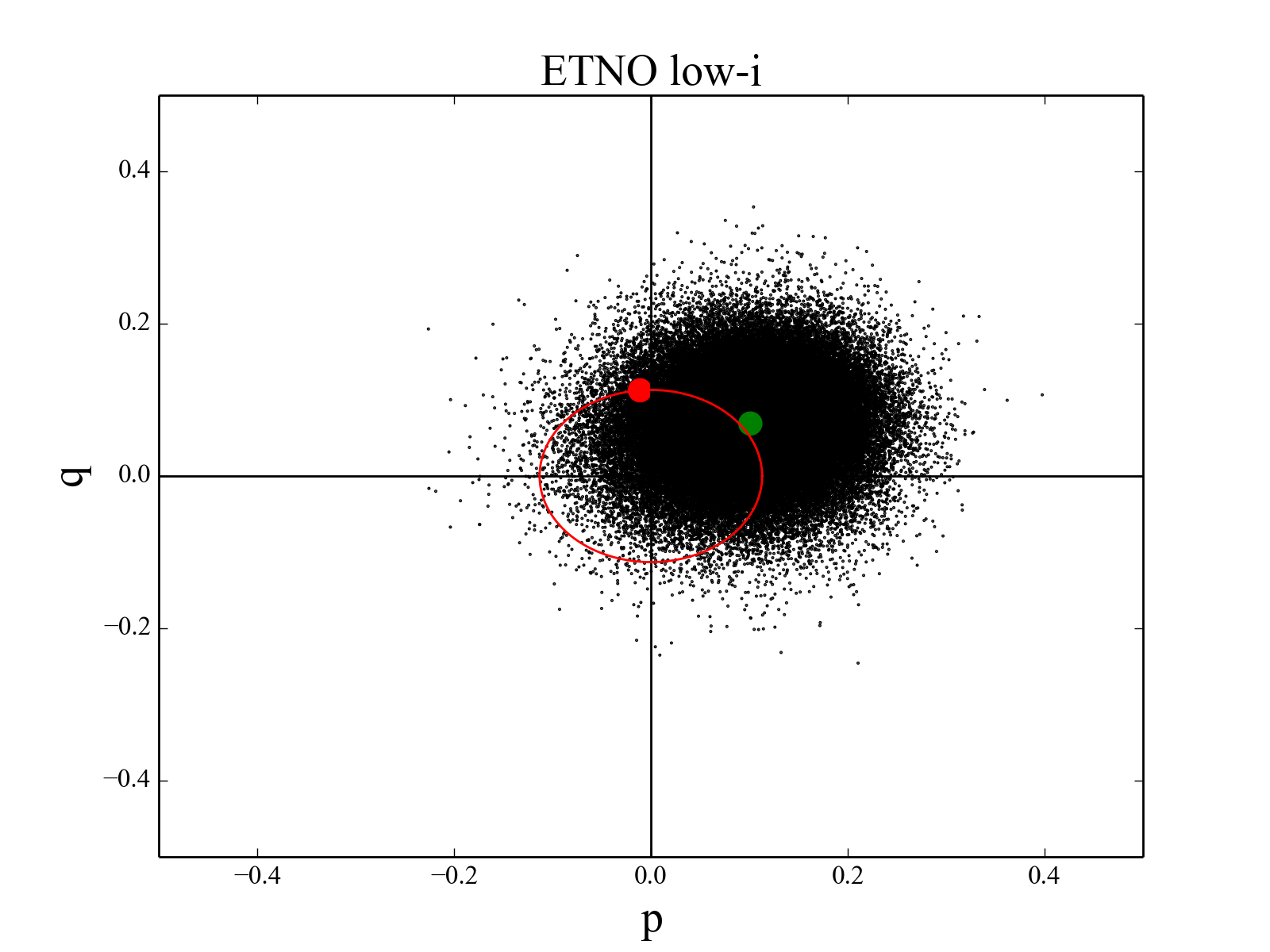}
	\qquad
	\includegraphics[width=0.4\textwidth]{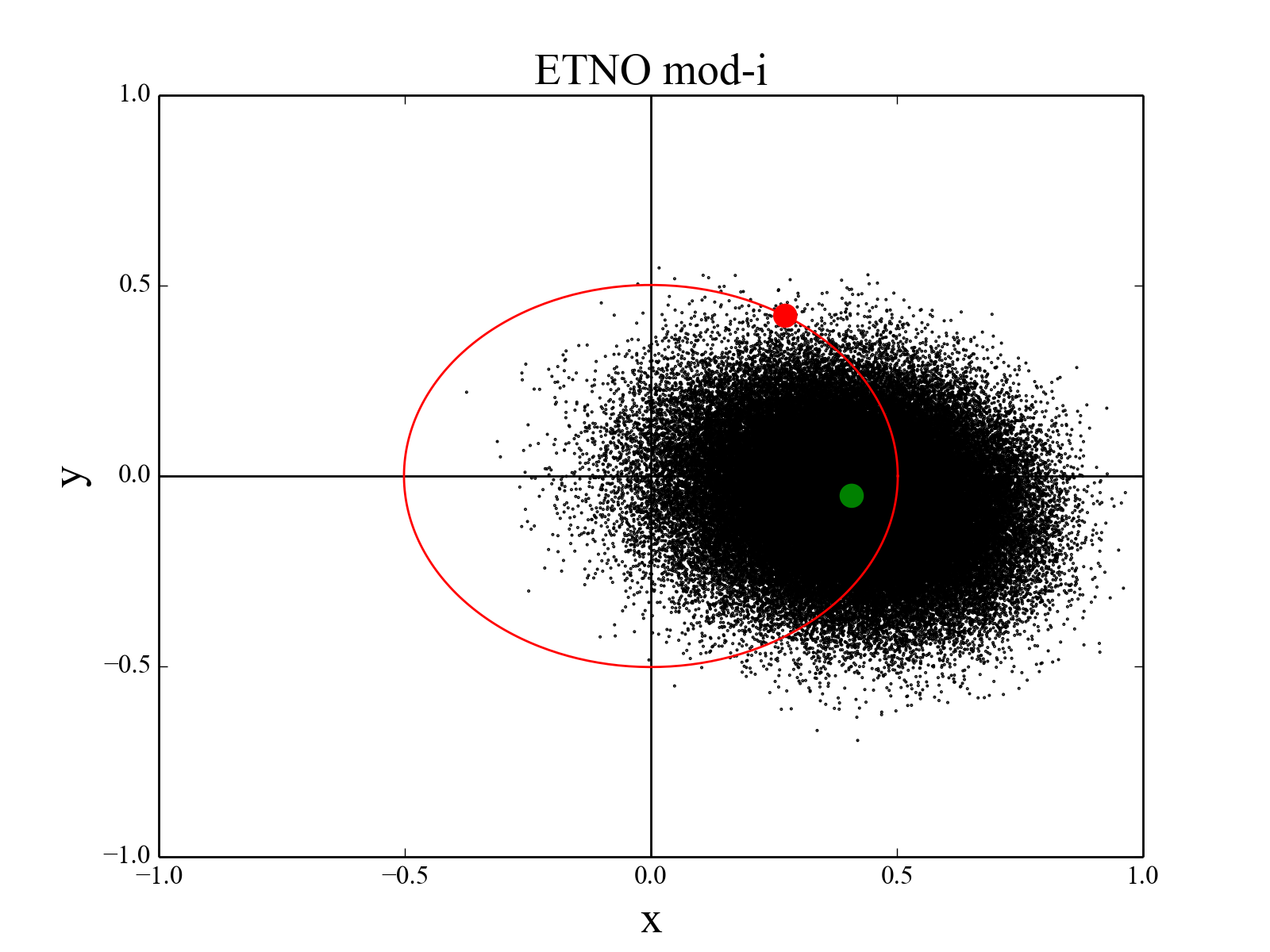}
	\includegraphics[width=0.4\textwidth]{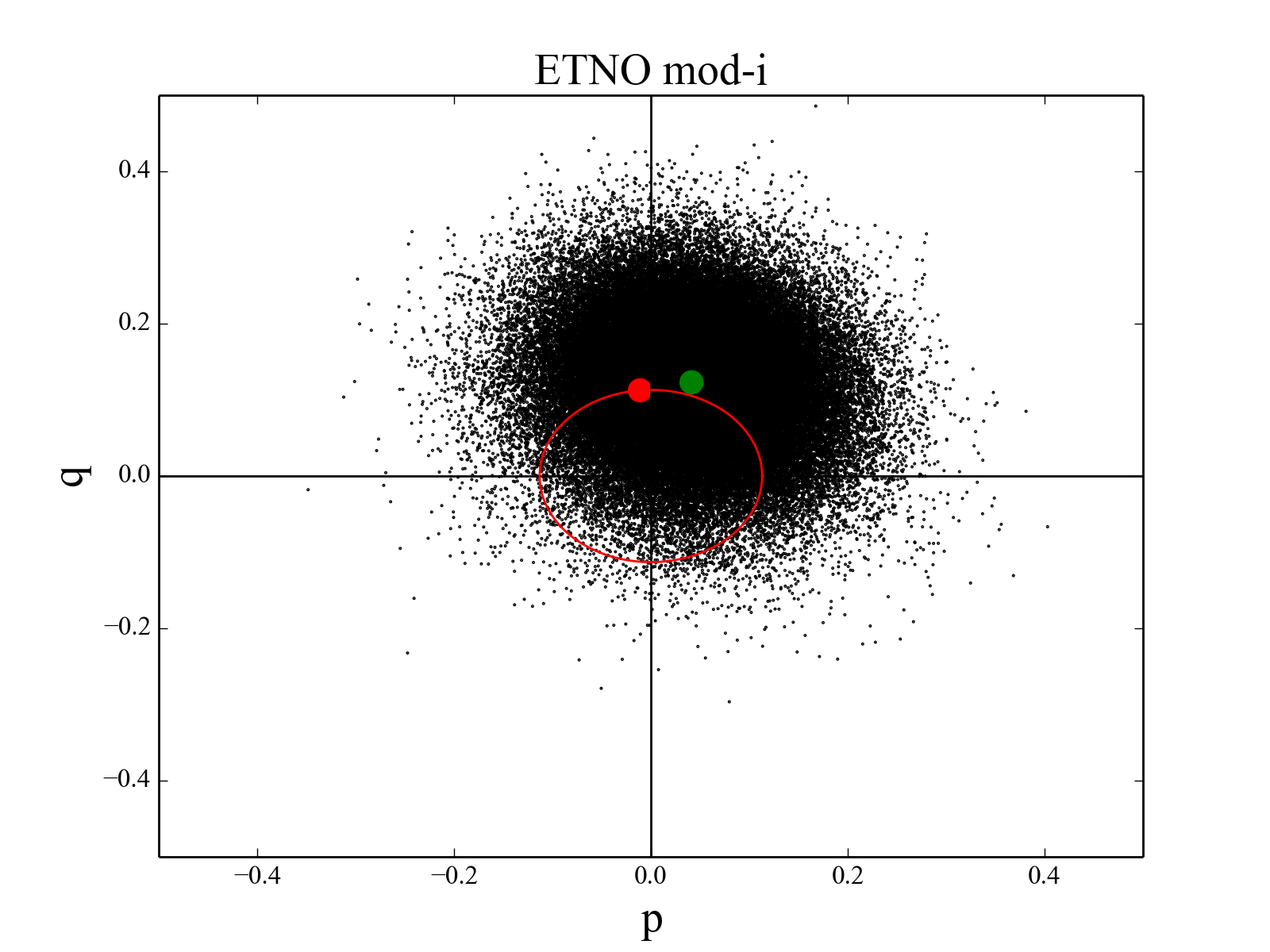}
	\qquad
	\includegraphics[width=0.4\textwidth]{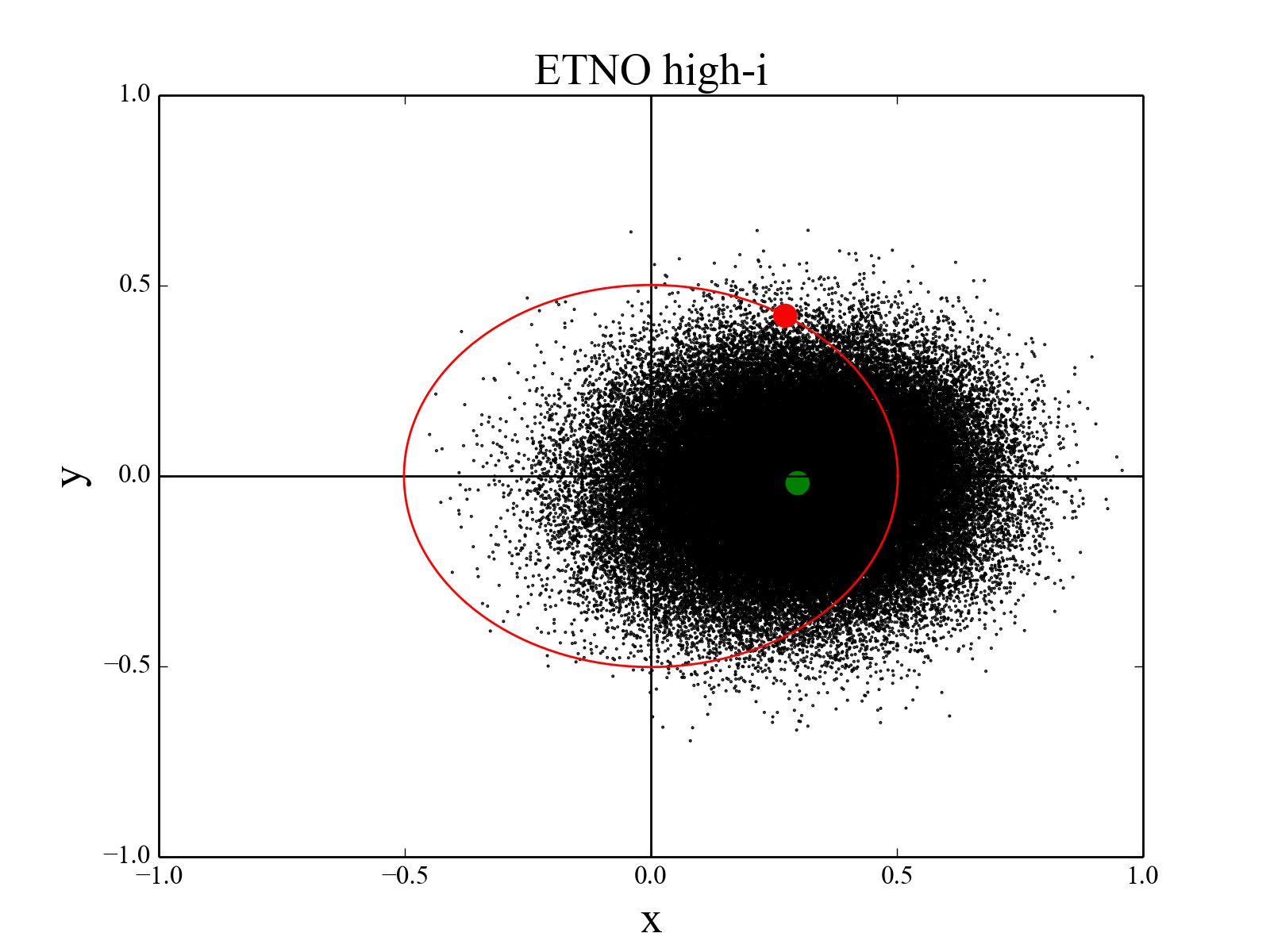}
	\includegraphics[width=0.4\textwidth]{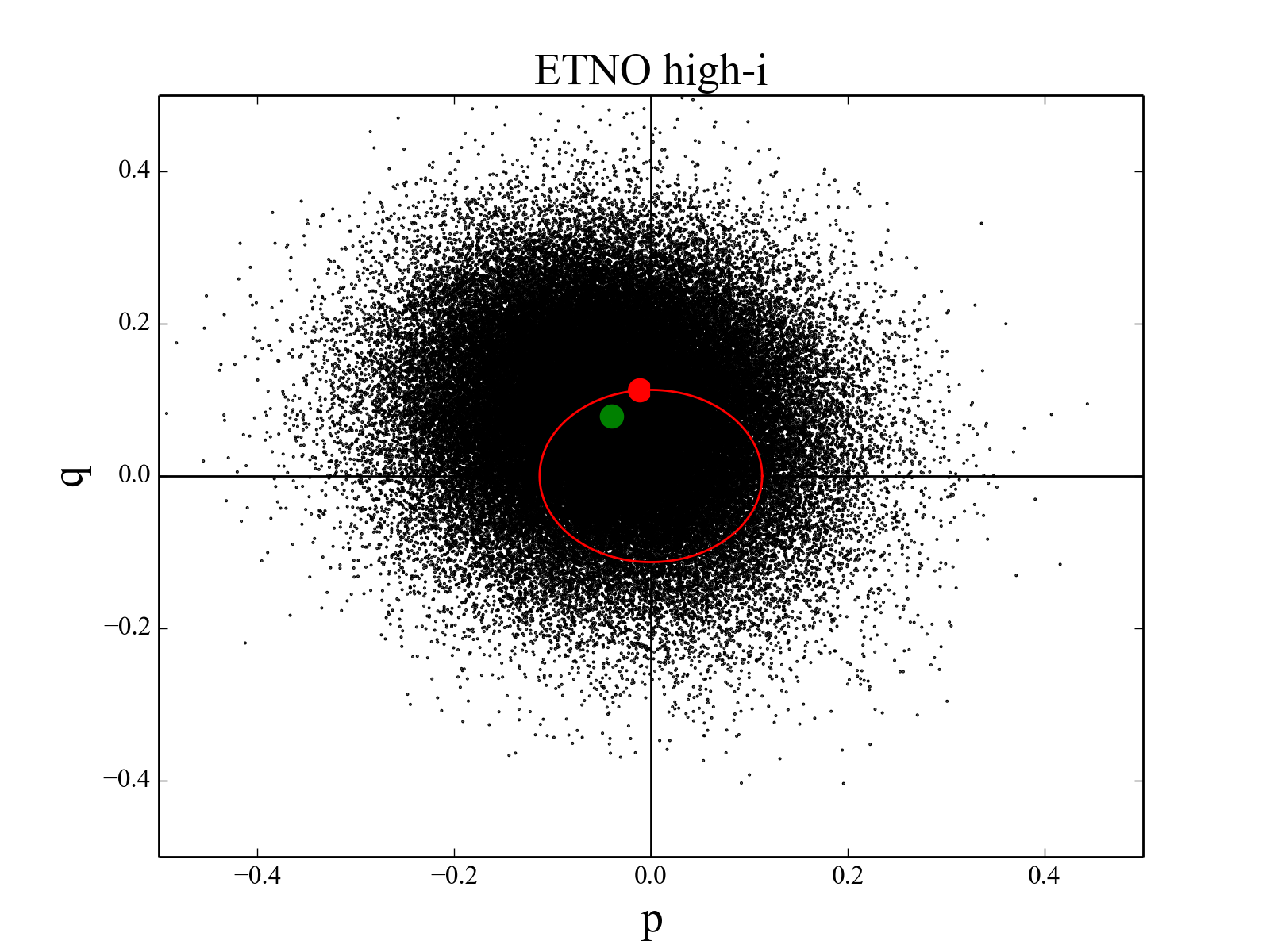}
	\caption{Same as figure \ref{fig:brown19} except, here, we randomly simulate 17 detections  from each of our ETNO, P9b simulations (calculating detection bias in the same manner).  Here, we find that the observed clustering (red point and circle) falls in line with our average, simulated clustering (66$\%$ and 69$\%$ of values falling outside the red circle in $x/y$ and $p/q$ space for the low-i batch, 20$\%$ and 67$\%$ for the mod-i set and 9$\%$ and 65$\%$ for the high-i simulations).}
	\label{fig:etno_xy}
\end{figure*}

\begin{figure*}
	\centering
	\includegraphics[width=0.4\textwidth]{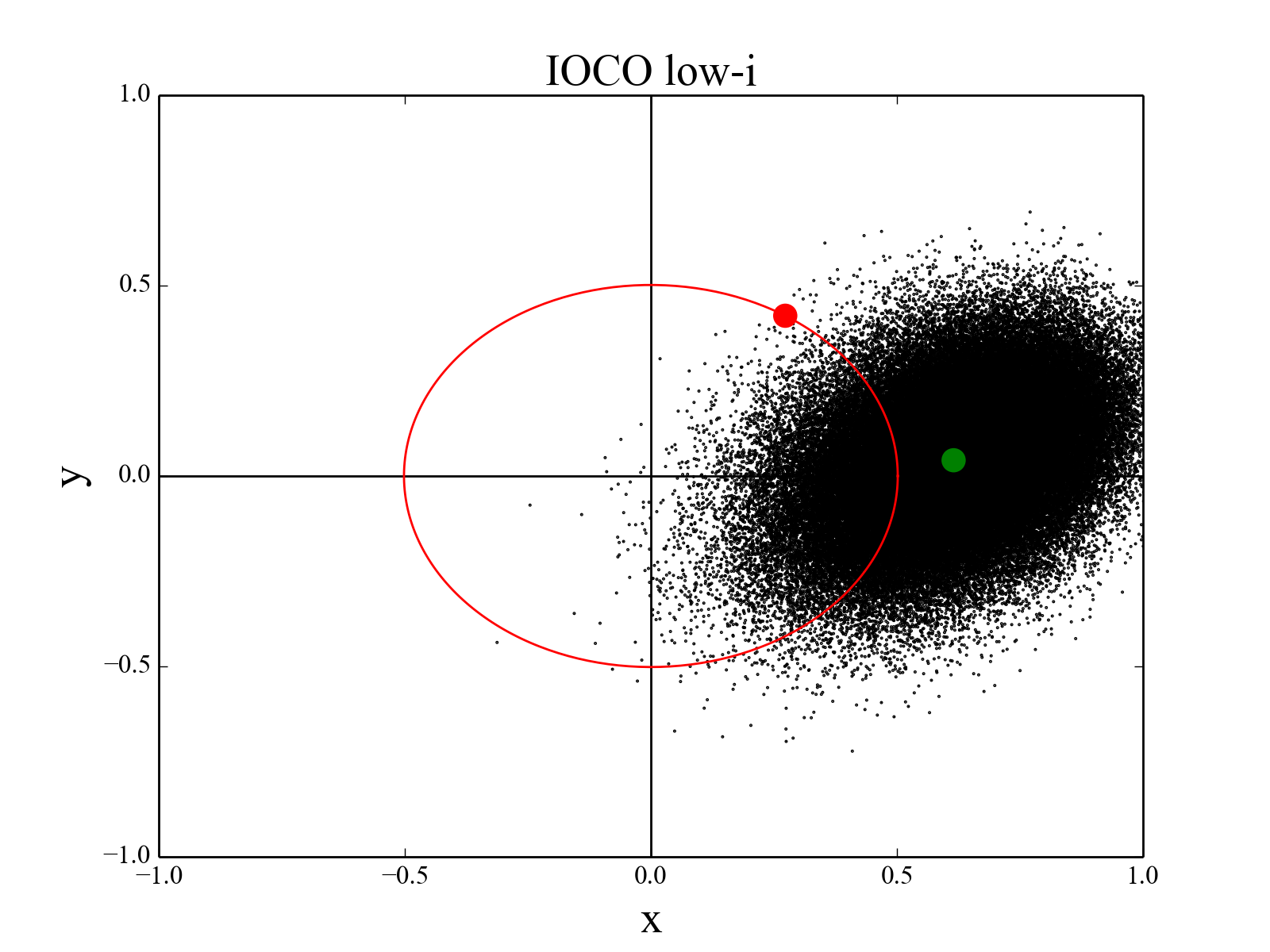}
	\includegraphics[width=0.4\textwidth]{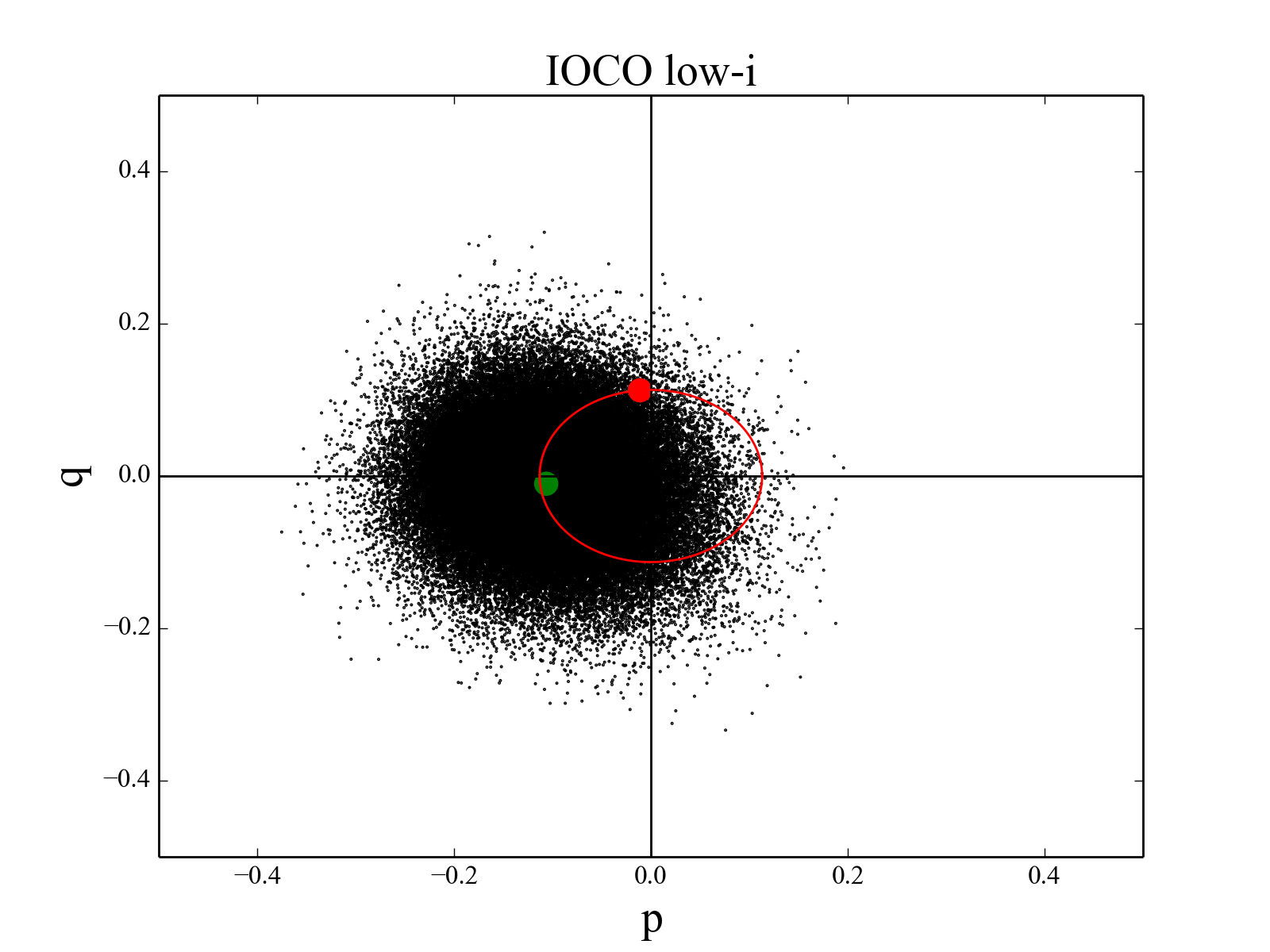}
	\qquad
	\includegraphics[width=0.4\textwidth]{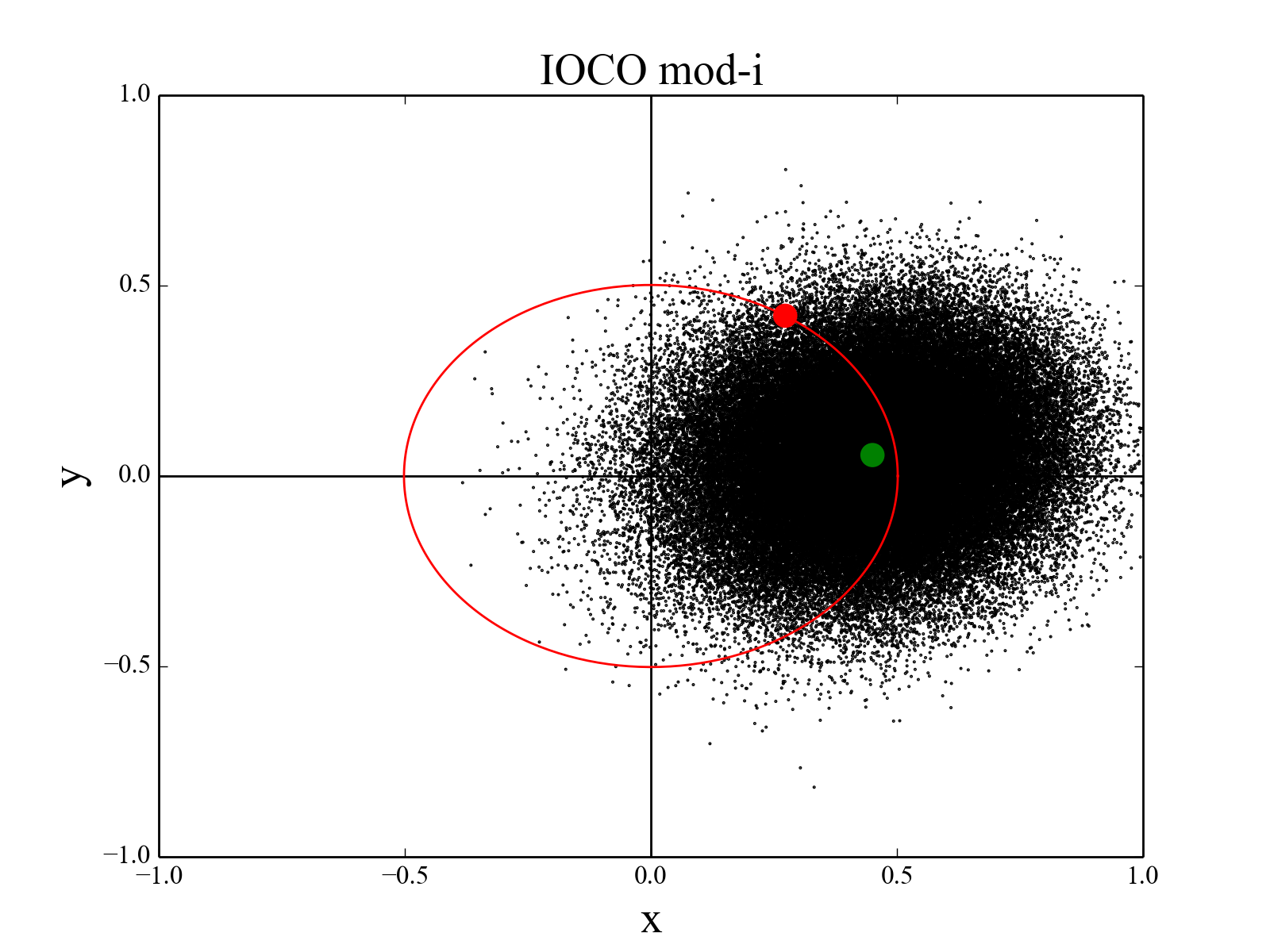}
	\includegraphics[width=0.4\textwidth]{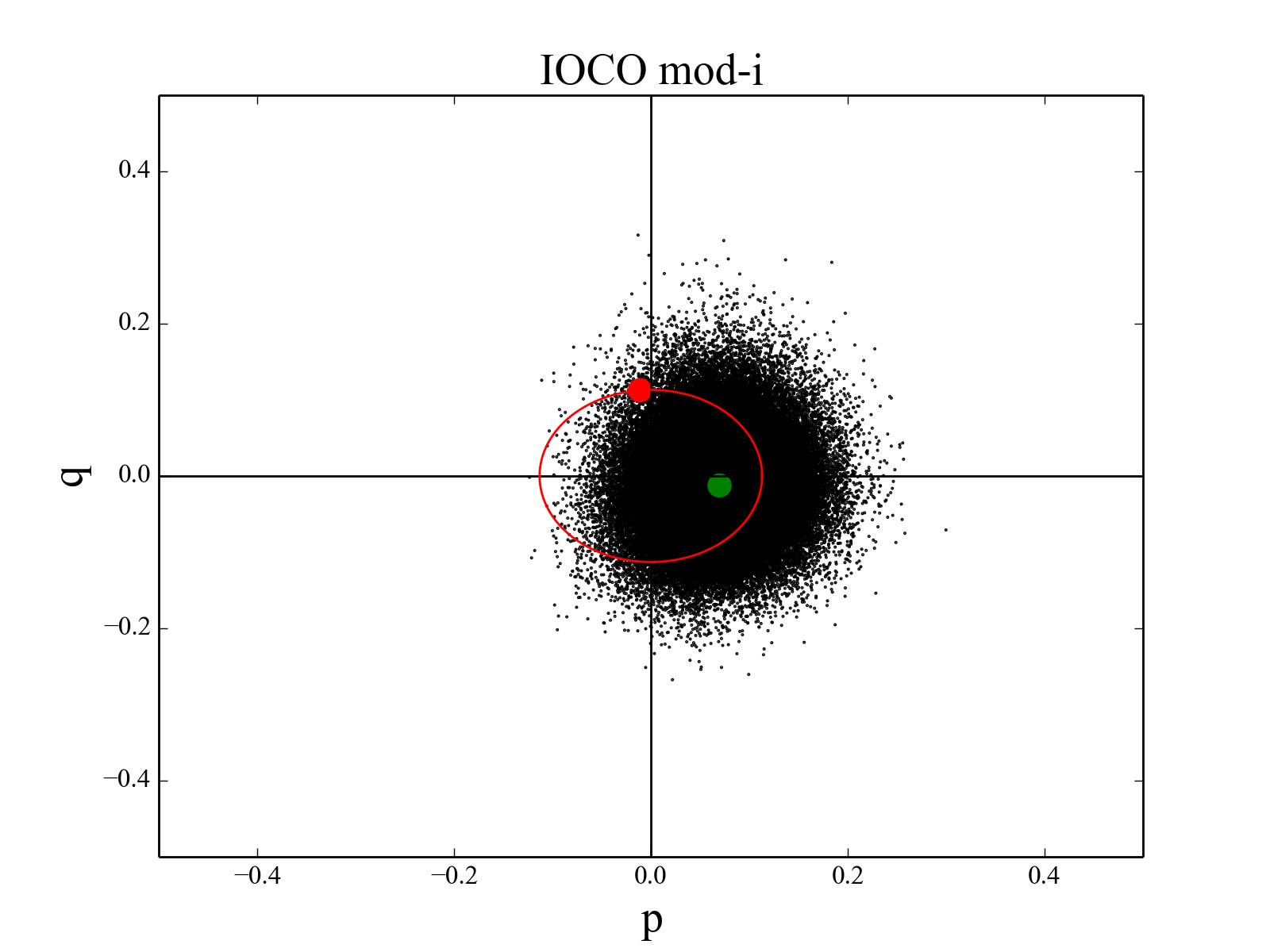}
	\qquad
	\includegraphics[width=0.4\textwidth]{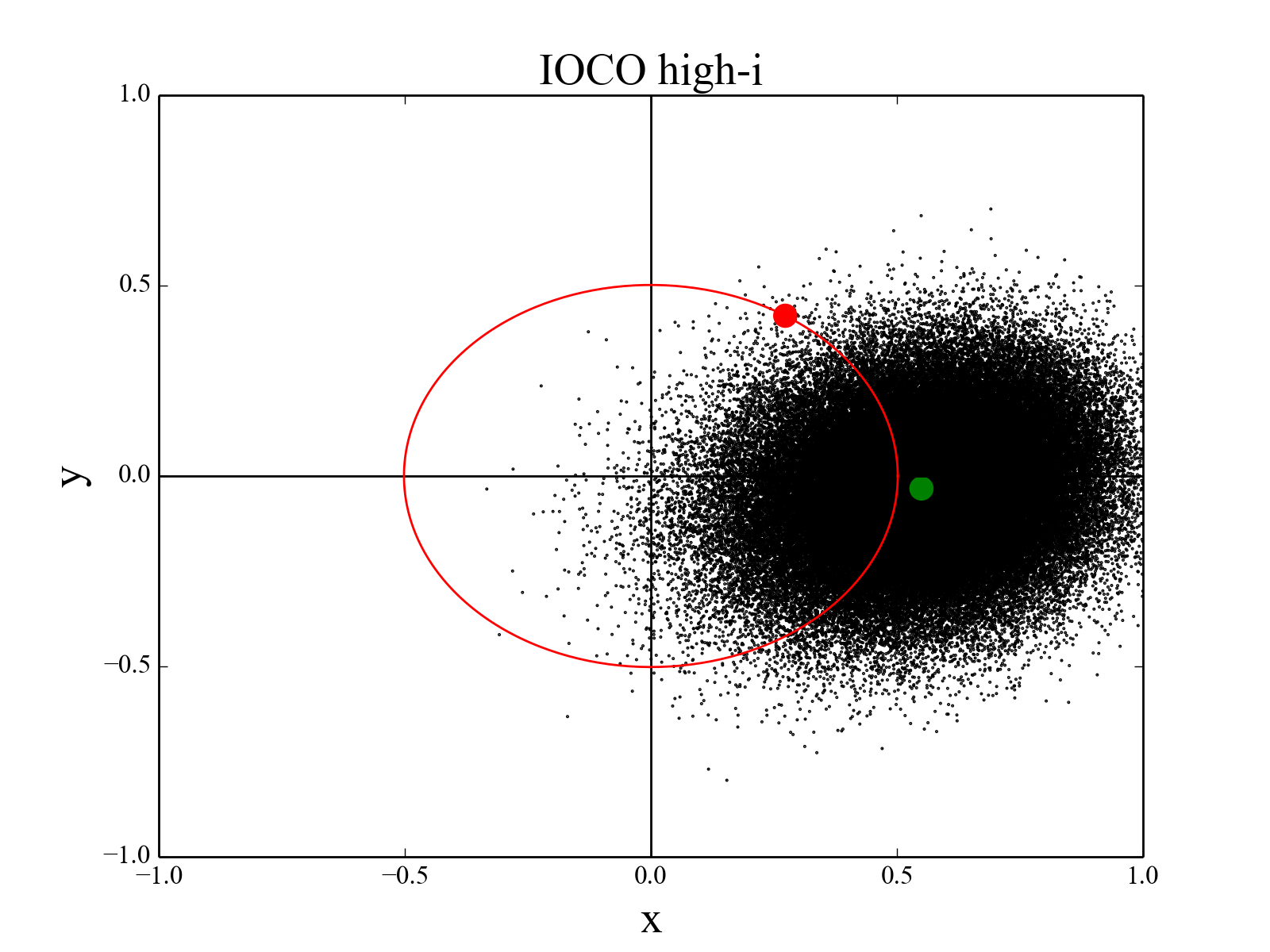}
	\includegraphics[width=0.4\textwidth]{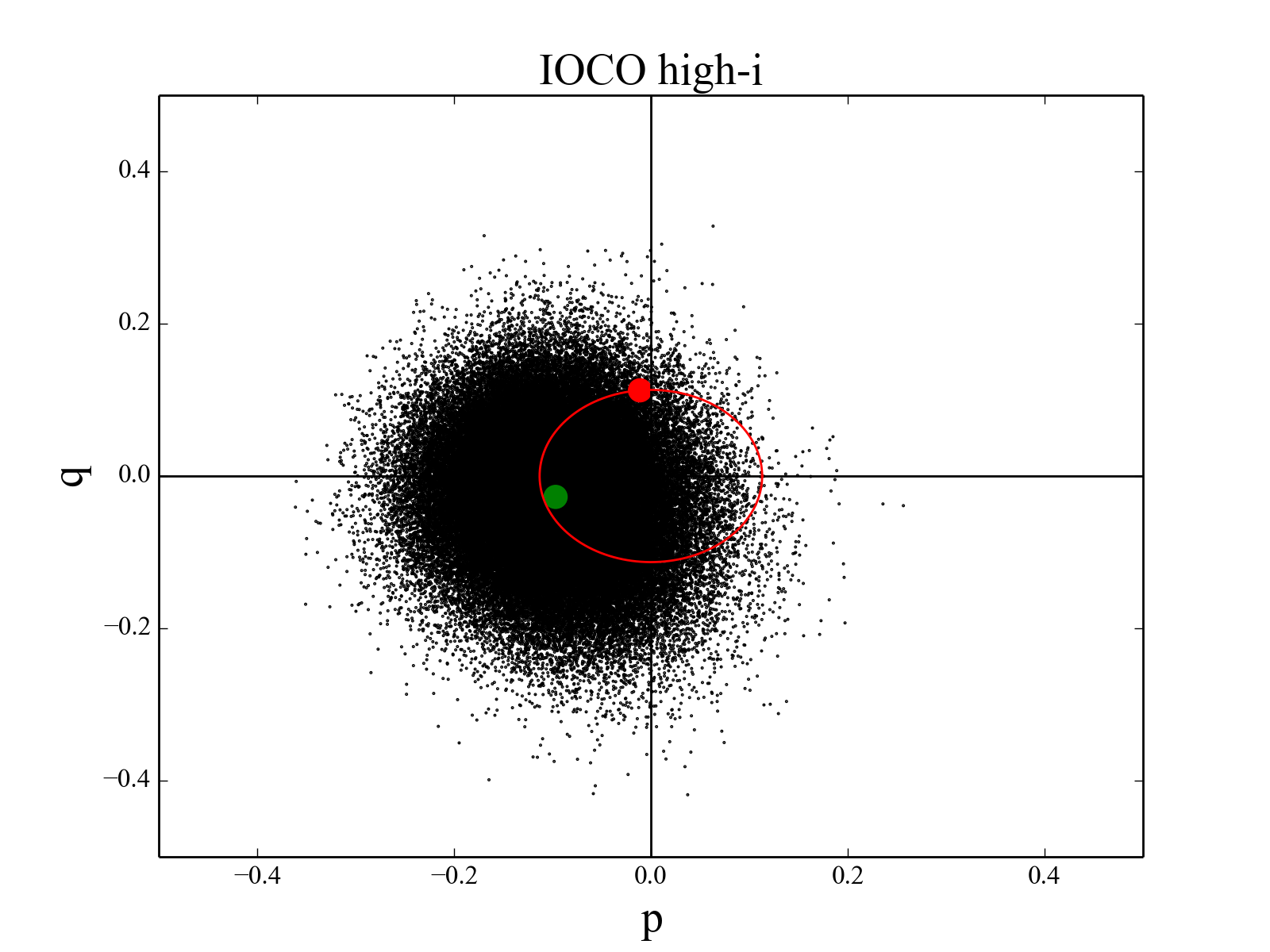}	
	\caption{Same as figure \ref{fig:etno_xy} except for our IOCO, P9b simulations.  Here, we find that the observed clustering (red point and circle) falls in line with our average, simulated clustering (67$\%$ and 61$\%$ of values falling outside the red circle in $x/y$ and $p/q$ space for the low-i batch, 56$\%$ and 40$\%$ for the mod-i set and 52$\%$ and 65$\%$ for the high-i simulations).}
	\label{fig:ioco_xy}
\end{figure*}

Figures \ref{fig:etno_xy} and \ref{fig:ioco_xy} plot this same test for our simulations designed to study bulk populations of ETNOs and IOCOs, respectively.  To generate these plots, we randomly draw 17 orbits from our de-biased population of simulated test particles \citep[here, we define an object as ``detectable'' if it has $q<$100 au after 4 Gyr, e.g.:][see further discussion in section \ref{sect:loss}]{sheppard19}.  While our P9b simulations consistently outperform their P9a counterparts, as expected, both sets of initial conditions are successful at generating clustered systems of distant orbits \citep{batygin16,batygin19_rev}.  However, it is important to point out the significant range of possible clusterings that can be generated when selecting just 17 orbits (thus, the noticeable overlap between figure \ref{fig:brown19} and figures \ref{fig:etno_xy} and \ref{fig:ioco_xy}).  This speaks to the problem of small number statistics that has become an intense topic of debate in the literature.  Thus, while the observed clustering is inconsistent with having originated from a uniform distribution of orbits at the $\lesssim$1$\%$ level \citep[figure \ref{fig:brown19},][]{batygin16,brown19}, it is not possible to detect the \textit{difference} between a uniform distribution and a Planet Nine generated distribution at greater than the 1$\sigma$ level.  To collapse these uncertainties and achieve a 2$\sigma$ distinction between our model generated orbital clusterings and a random sample of uniformly distributed objects requires $\gtrsim$100 detections.  In other words, a small number of detected objects carry a greater inherent bias, which manifests in figures \ref{fig:brown19}, \ref{fig:etno_xy} and \ref{fig:ioco_xy} as a large range of deviations from the true underlying clustering (green point).  \citet{kaib19} reached a similar conclusion when studying the genesis of scattered TNOs as constrained by the OSSOS dataset.  In figure 6 of that work, the authors estimate that $\gtrsim$60 detections of high-$q$ objects would be required to distinguish a Planet Nine model from the uniform case. 

The precise \textit{degree} of deviation from the origin for our clustered ETNO and IOCO orbits plotted in figures \ref{fig:etno_xy} and \ref{fig:ioco_xy} is also a function of the underlying distribution of eccentricities and inclinations.  Because the magnitudes of vectors in $x/y$ and $p/q$ space scale with eccentricity and inclination, respectively, the range of possible detected clusterings (black points) scale likewise.  This manifests most significantly in $p/q$ space for our simulations, given the fact that we test different inclination distributions.  This has the obvious effect of increasing the degree of scatter in systems that possess more high inclination objects (right panels of figures \ref{fig:etno_xy} and \ref{fig:ioco_xy}).  Our plotted ``detections'' are bias-informed such that low inclination objects that spend more time at the ecliptic are more likely to be detected.  In spite of this correction, the effect of greater scatter with increasing $\sigma_{i}$ persists because high-$i$ particles still cross the ecliptic and can thus still be ``detected,'' albeit at a lower frequency.

In table \ref{table:big_tab}, we provide relevant statistics for our various P9b simulation sets.  It is important to note that these extremely distant objects' orbits are highly perturbed by the galactic tide and the giant planets over Gyr-timescales.  External forces cause object's perihelia to fluctuate by tens of au, often driving them on to orbits where they interact weakly with the giant planets and can be ejected \citep[e.g.:][]{sheppard19}. The extremely low fractions ($\sim$10-25$\%$) of test particles and object clones that remain detectable after 4 Gyr speaks to how difficult it is to study this region of the solar system with such a small sample of observed objects.

There is a clear difference between our ETNO and IOCO sets in terms of the stability of test particles and clustering in $\varpi$ with respect to inclination.  In general, longitude of perihelion clustering is most efficient in our lowest inclination sets (also visible in figures \ref{fig:etno_xy} and \ref{fig:ioco_xy}).  This is a result of high inclination dynamics randomizing objects' $\varpi$ and $\Omega$ as the secular shepherding weakens.  We direct the reader to equation 8 of \citet{batygin19_rev} for a full derivation and explanation.

In general, objects become decoupled from Planet 9 when their perihelion is driven in to the planetary regime.  This can be understood in terms of orbital precession induced by the Sun's artificial $J_{2}$ component \citep[for a full derivation see, for example:][]{murray_dermott}:
\begin{equation}
	J_{2} = \frac{1}{2}\sum_{i=5}^{8}\frac{m_{i}a_{i}^{2}}{M_{\odot}R_{\odot}^{2}}
\end{equation}
The longitude and perihelion and nodal precessions driven by $J_{2}$ are \citep[e.g.:][]{brasser06}:
\begin{equation}
	\dot{\varpi} = \frac{3n}{4}\bigg(\frac{R_{\odot}}{a}\bigg)^{2}J_{2}\bigg(\frac{5\cos^{2}{i}-1}{(1-e^{2})^{2}}\bigg)
\end{equation}
\begin{equation}
	\dot{\Omega} = -\frac{3n}{2}\bigg(\frac{R_{\odot}}{a}\bigg)^{2}J_{2}\bigg(\frac{\cos{i}}{(1-e^{2})^{2}}\bigg)
\end{equation}

Once the perihelion of an object is lowered into the planetary regime, stronger $J_{2}$ perturbations can result in its value of $\varpi$ becoming decoupled from Planet Nine.  This effect leads to correspondingly lower fractions of $N_{\varpi}$/$N_{detect}$ for higher inclination objects in table \ref{table:big_tab}.   This process is more efficient for ETNOs (56$\%$ of low-$i$ batch clustered in $\varpi$ compared with 38$\%$ of high-$i$) than for IOCOs since they are already significantly detached from the giant planet system (56$\%$ of low-$i$ batch clustered in $\varpi$ compared with 51$\%$ of high-$i$).  In the same manner, only 12$\%$ of our ETNO high-$i$ set, and 6$\%$ of the low-$i$ set still have q $<$ 100 au after 4 Gyr of evolution.  Consistent with previous authors we conclude that, as a broad trend, the strength of longitude of perihelia shepherding weakens with increasing inclination and decreasing perihelion distance.

\begin{table}
	\centering
	\begin{tabular}{c c c c c}
	\hline
	Run & $q_{o}$ & $N_{detect}$/$N_{o}$ & $N_{\varpi}$/$N_{detect}$ & $N_{\Omega}$/$N_{detect}$ \\
	\hline
	Control & 40-50 & 0.91 & 0.25 & 0.29 \\
 	IOCO,low-$i$ & 40-50 & 0.11 & 0.56 & 0.37  \\
 	IOCO,mod-$i$ & 40-50 & 0.19 & 0.54 & 0.41  \\
 	IOCO,high-$i$ & 40-50 & 0.19 & 0.51 & 0.41  \\
 	VP113 & 80.3 & 0.24 & 0.51 & 0.42 \\
 	Sedna & 76.0 & 0.10 & 0.51 & 0.38 \\
 	TG387 & 65.0 & 0.17 & 0.65 & 0.55  \\
	ETNO,low-$i$ & 50-100 & 0.23 & 0.56 & 0.52 \\
 	ETNO,mod-$i$ & 50-100 & 0.06 & 0.47 & 0.45  \\
 	ETNO,high-$i$ & 50-100 & 0.12 & 0.38 & 0.39  \\
 	SY99 & 49.9 & 0.14 & 0.78 & 0.63  \\
 	GB174 & 48.8 & 0.23 & 0.46 & 0.44\\
 	SR349 & 47.6 & 0.06 & 0.40 & 0.30  \\
 	VN112 & 47.3 & 0.20 & 0.33 & 0.32 \\
 	RX245 & 45.5 & 0.16 & 0.60 & 0.49  \\
 	\hline
	\end{tabular}
	\caption{Summary of simulation results for our P9b configurations.  The columns are as follows: (1) The simulation set, (2) The initial pericenter, (3) the fraction of objects still ``detectable'' (here defined as $q<$100 au) after 4 Gyr, (3) the fraction of detectable objects anti-aligned ($\pm$ 45$^{\circ}$) with Planet Nine in $\varpi$ after 4 Gyr (for our Control simulations this is the fraction of objects within $\pm$ 45$^{\circ}$ of the circular mean in $\varpi$)  and (5) the fraction of detectable objects anti-aligned ($\pm$ 45$^{\circ}$) with Planet Nine in $\Omega$ after 4 Gyr.}
	\label{table:big_tab}
\end{table}

\subsection{Significance of each known object}
\label{sect:loss}

\begin{figure}
	\centering
	\includegraphics[width=0.5\textwidth]{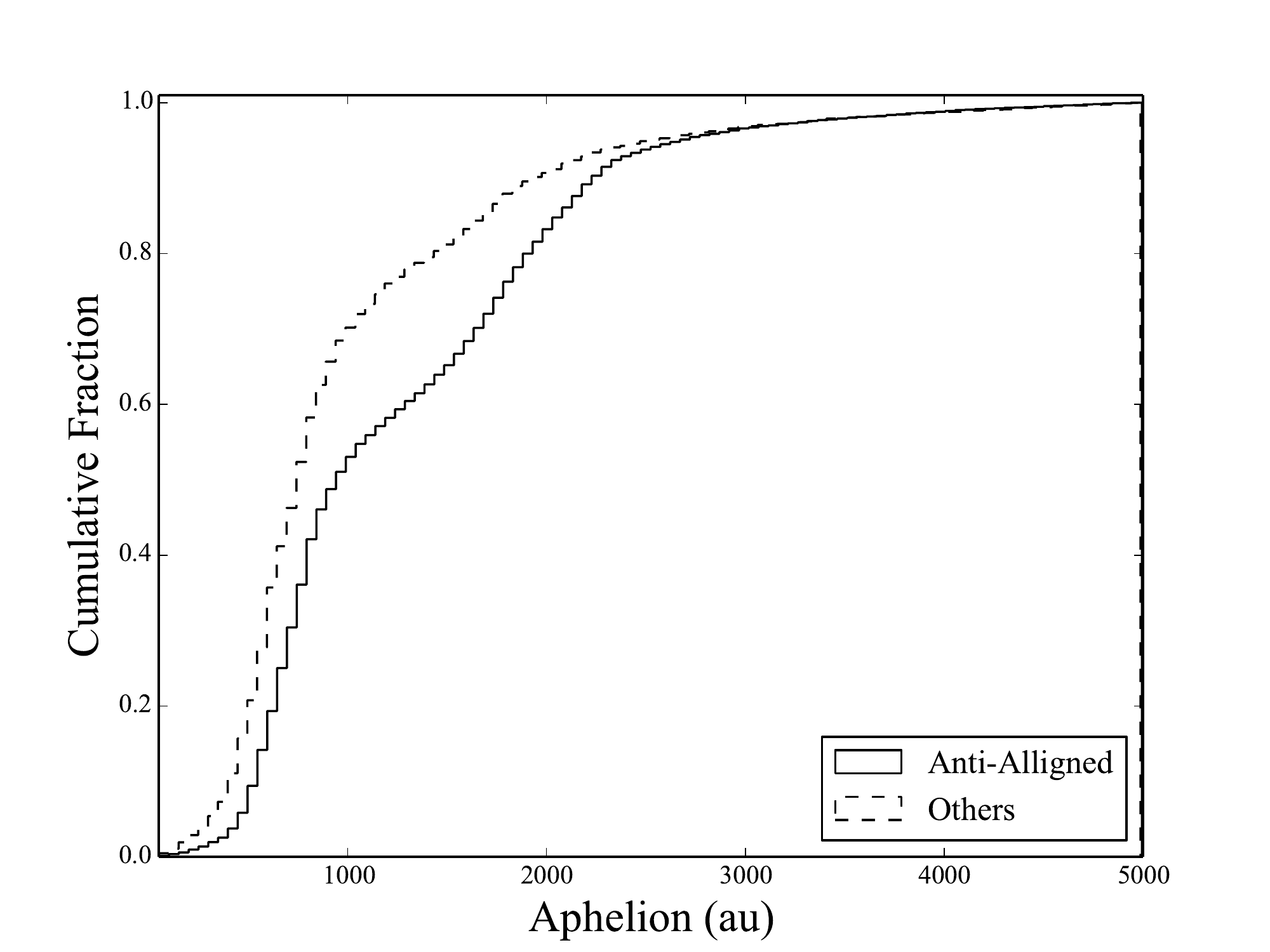}
	\caption{Cumulative fraction of test particle and clone aphelia for those objects anti-aligned with Planet Nine in longitudes of perihelia ($\pm$ 45$^{\circ}$, solid line), versus objects that are not (dashed line).}
	\label{fig:hist}
\end{figure}

In general, we find similar fractions of objects detectable ($\sim$10-25$\%$), and clustered in $\varpi$ ($\sim$50-65$\%$) and $\Omega$ ($\sim$35-50$\%$) after 4 Gyr from within our various sets of clone simulations.  However, we note extremely high levels of clustering for our surviving 2013 SY99 clones (78$\%$ clustered in $\varpi$), and relatively poor levels for 2014 SR349 and 2004 VN112 clones.  This is a result of Planet Nine's orbital shepherding having a higher efficiency for the most detached objects; or those that spend the greatest fractions of time at large distances away from the giant planets.  Thus it follows that, of the 5 ETNO clones we test (each with similar perihelia), SY99 has the largest aphelion distance ($>$1,400 au) and also anti-aligns most efficiently with Planet Nine.  Contrarily, SR349 and VN112 have aphelia close to 600 au.  We demonstrate this trend in figure \ref{fig:hist} by plotting the aphelia of objects (from all of our simulations which include Planet Nine) that are detectable and opposed in $\varpi$ with Planet Nine after 4 Gyr against the remainder of detectable particles.  This is the reason that the most probable orbits \citep[e.g.:][]{batygin19_rev} for the hypothetical planet place its longitude of perihelion almost exactly 180$^{\circ}$ in opposition of the detached TNOs with the most extreme aphelia (namely SY99 and TG387).

\subsection{A chance alignment}
\label{sect:inc}

\begin{figure}
	\centering
	\includegraphics[width=.5\textwidth]{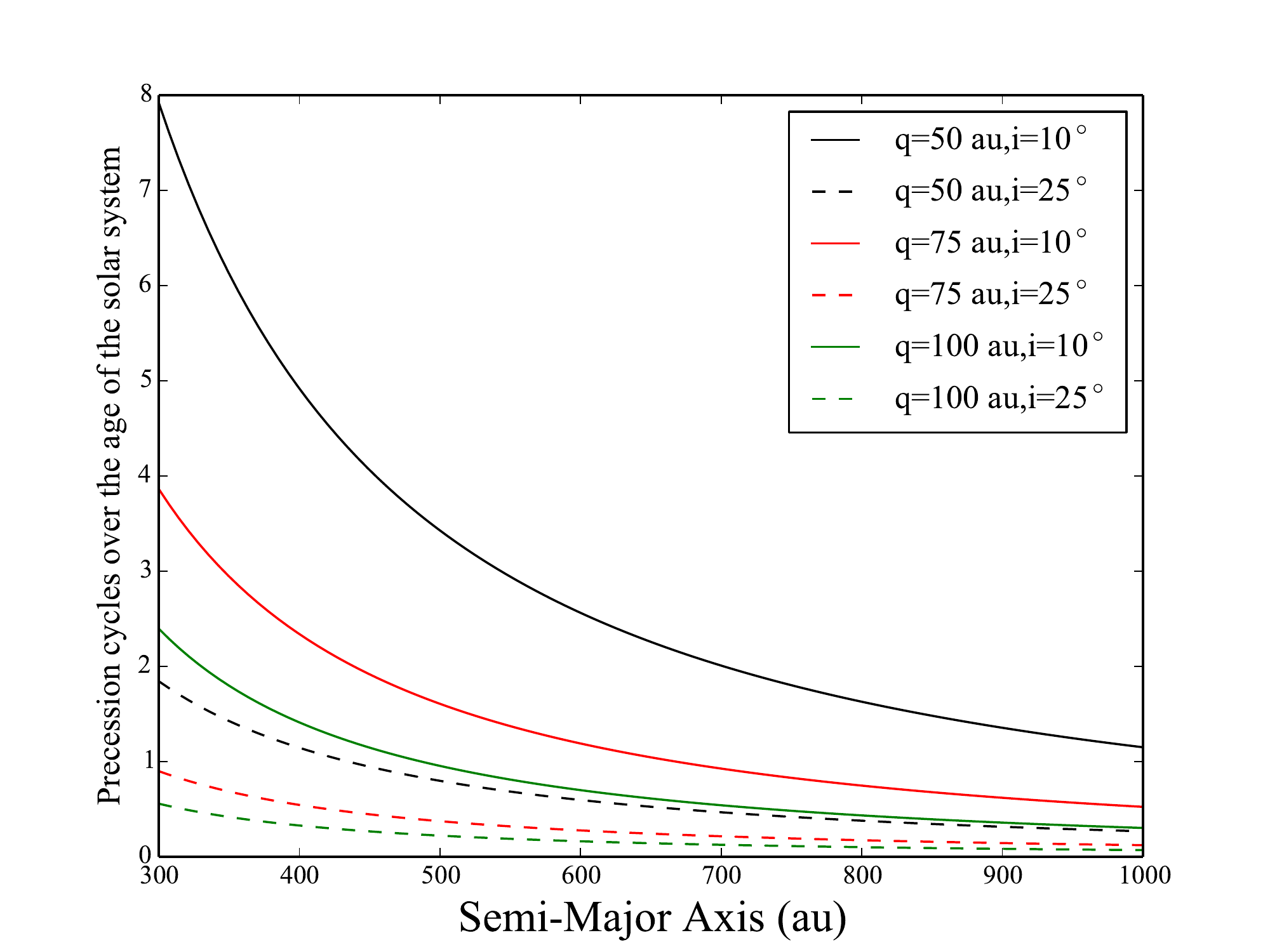}
	\caption{$J_{2}$ induced precession (cycles per 4.5 Gyr) of $\varpi$ in the distant solar system as a function of semi-major axis for different values of inclination (solid and dashed line styles) and perihelion (black, red and green line colors).}
	\label{fig:precession}
\end{figure}

Absent an external perturber, the orbits of ETNOs and IOCOs precess exceedingly slow as the result of perturbations from the giant planets.  Indeed, we would expect that, only under the influence of $J_{2}$, the longitudes of perihelia for most of the known objects in the region would traverse a full 360$^{\circ}$ only a few times over the life of the solar system (figure \ref{fig:precession}).  The most extreme objects \citep[for example Sedna:][]{brasser06} would only be expected to make $\sim$one precession in 4 Gyr.  

Our simulations (section \ref{sect:disc}) lead us to conclude that the possible levels of orbital clustering that might be observed from a Planet Nine generated distribution of TNOs and that of a uniform sample overlap significantly (because of the small number of known extreme objects).   Thus, while the observed degree of clustering is highly significant in spite of small number statistics and observational biases \citep{brown19}, its specific connection with a 5.0 $M_{\oplus}$ planet orbiting around 500 au is not.  However, there is an obvious absence of equally robust explanations in the literature \citep{batygin19_rev}.  While a chance or coincidental alignment might seem compelling given the minute degree of $J_{2}$ induced precession in the distant solar system (figure \ref{fig:precession}), the probability of such an event and the corresponding dispersion timescale for a system of spontaneously clustered orbits has yet to be characterized.  As such, we conclude this manuscript by commenting on the slow dispersion timescale for clustered orbits in the distant solar system.

For this phase of our study, we select the $Mercury6$ Hybrid integrator \citep{chambers99} and utilize a 180 day time-step ($\sim$4$\%$ of Jupiter's orbital period).  In each simulation, we include the solar system's four giant planets and 50, Pluto-massed ETNOs (here defined as 300 $<a<$ 1,000 au and 45 $<q<$ 80 au; loosely based off the sample of detected ETNOs and IOCOs; e.g.: table \ref{table:ics2}).  ETNOs begin with identical values of $\Omega$ and $\varpi$, randomly selected semi-major axes, perihelia, and mean anomaly.  All simulation particles interact gravitationally with one another.  In 25 simulations, we analyze ETNOs with co-planar inclinations ($i=$ 0$^{\circ}$).  In an additional set of 25 integrations, we assign inclinations in accordance with equation \ref{eqn:fi} ($\sigma_{i}=$ 10.0$^{\circ}$, akin to our mod-$i$ simulations of section \ref{sect:disc}).  Note also that these simulations do not include algorithms that account for the galactic tide or stellar encounters, as we seek to minimize unnecessary particle loss. 

\begin{figure}
	\centering
	\includegraphics[width=.5\textwidth]{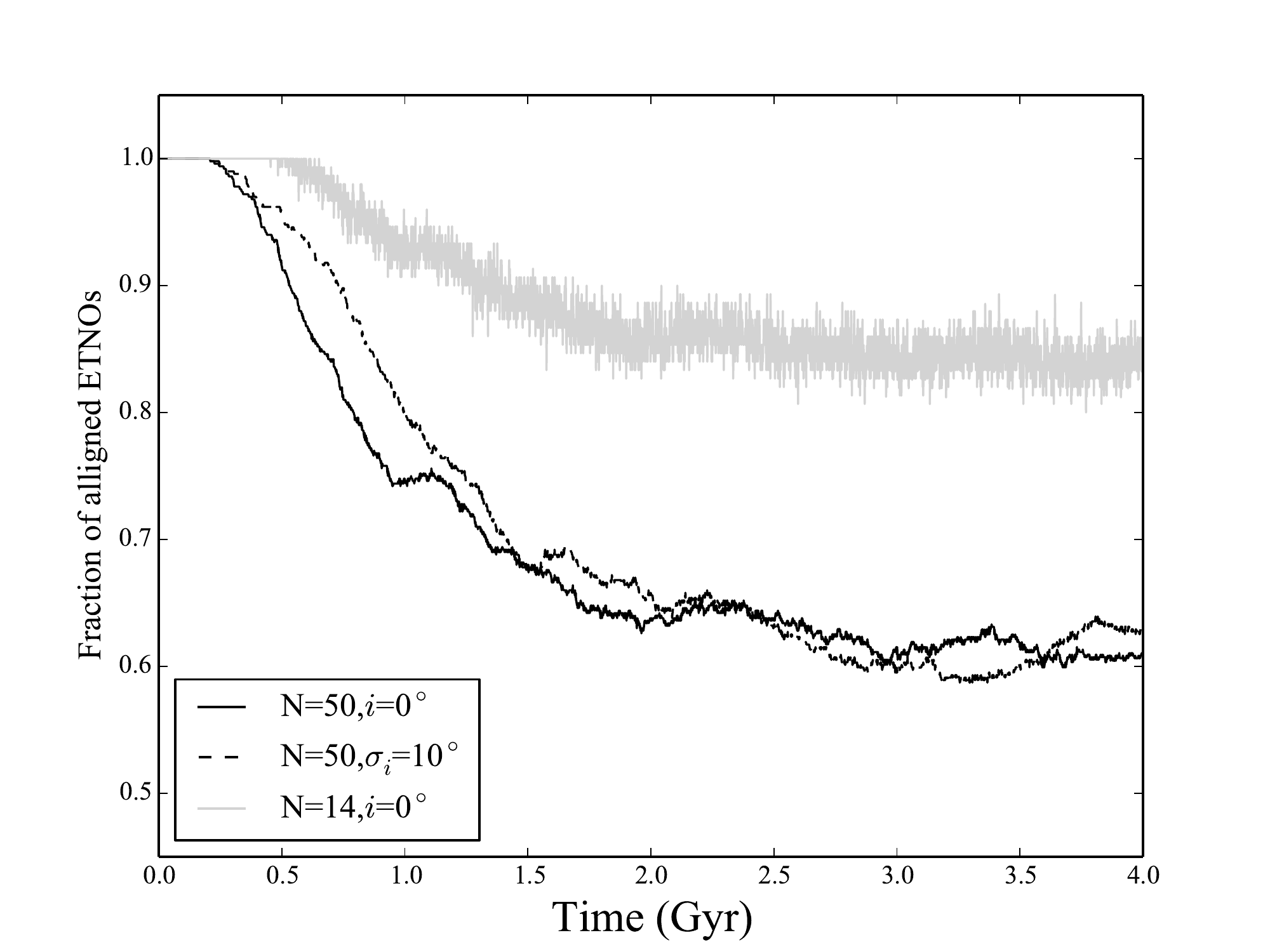}
	\caption{Evolution of a system of ETNOs that begin with identical values of $\varpi$ and $\Omega$.  At each simulation time output, we select the value of $\varpi$ ($\varpi_{test}$) that maximizes the fraction of objects with $\varpi$ within $\pm$90$^{\circ}$ of $\varpi_{test}$.}
	\label{fig:360}
\end{figure}

Figure \ref{fig:360} plots the average dispersal of longitudes of perihelia in our 50 simulations.  At 1 Myr intervals, we calculate the angle ($\varpi_{test}$) that maximizes the ratio of objects with $\varpi$ within $\pm$90$^{\circ}$ of $\varpi_{test}$ to those anti-aligned.  Consistent with figure \ref{fig:precession}, our higher inclination population of TNOs remain ``aligned'' for longer ($\sim$250 Myr) as they are less affected by the giant planets.  It is interesting that, in both cases, the orbital alignment persists for over a Gyr, requiring $\sim$2 Gyr to reach an equilibrium value of $\sim$60$\%$ aligned objects.   Furthermore, the deviation from a 50$\%$ equilibrium value is intriguing in and of itself.  This is a direct representation of the effects of small number statistics on a binary division of angles where $\varpi$ (or $\Omega$ for clustering in orbital pole) is taken as a free parameter (e.g.: circular statistics).  This speaks to the fundamental assumption of attempts to de-bias the population of ETNO and IOCO detections: the pre-condition that, in the absence of external perturber, the objects' orbits would be drawn from a uniform distribution.  Or, rather in our case, supposing that the detection of all KBOs with respect to $\varpi$ or $\Omega$ is representative of the probability of detecting any extreme object with a given value of $\varpi$ or $\Omega$.  As shown in figure $\ref{fig:360}$, when the underlying population of objects is finite, and $\varpi$ is taken as a free parameter, the underlying distribution of angles is inherently biased and non-uniform (this idea is analogous to the observed scatter in our iterative selection of 17 random orbits from within our Control set in figure \ref{fig:brown19}).  This is easily exploited by selecting a subset of just 14 objects (grey line).  With only 14, randomly drawn angles in the dataset, it is easy to find an angle that a large fraction (equilibrium value close to $80\%$) $appear$ to cluster around.  This argument would obviously not hold for the classical Kuiper Belt as a whole, as it is fairly well characterized observationally.  However, the underlying population from which the observed population of detached ETNOs and IOCOs originate is $not$ the entire ETNO and IOCO constituency or bodies.  Rather, it is the subset of objects (each with orbital periods of order $\sim10^{4}$ yr) that are close to perihelion $now$.  Such a collection of objects is finite in number, and thus likely to be concentrated around some orbital elemental angles when ``alignment angle'' is taken as a free parameter.

\section{Conclusions}

In this paper, we presented a large suite of dynamical simulations designed to study the stability of detached, distant objects (Extreme Trans-Neptunian Objects: ETNOs, and Inner Oort Cloud Objects: IOCOs) in the presence of the hypothetical Planet Nine \citep{batygin16,batygin19_rev}.  Our numerical simulations included algorithms that approximate the effects of the passing stars \citep{fernandez00,rickman08,fabo15} and the galactic tide \citep{heisler86,dones04,kaib11}.  By and large, our simulations find that $\sim$5-20$\%$ of extreme objects are still ``detectable'' after 4 Gyr; $\sim$50-70$\%$ of which have orbital shapes anti-aligned with Planet Nine.  In general, objects least affected by the giant planets (those with large aphelia) are best shepherded by the external perturber.  Thus, we find that clones of the most extreme objects 2013 SY99 and 2005 TG387 anti-align most efficiently with Planet Nine, while 2014 SR349 and 2004 VN112 clones are least likely to remain stable in anti-aligned orbits for 4 Gyr.  Moreover, we compare the apparent orbital clustering of 17 objects drawn at random from a uniform distribution with that of our simulated objects (accounting for the observational biases).  While we confirm that the 17 extreme objects used to infer the existence of Planet Nine are inconsistent with having been drawn from a uniform distribution at the $\lesssim$1$\%$ level, we cannot distinguish between the two distributions with any reasonable statistical significance.  That is to say, the expected Planet Nine-induced orbital clustering in the distant solar system (by a 5.0 $M_{\oplus}$ planet at $\sim$500 au) is not strong enough to distinguish from the null case using only 17 objects.

We perform an additional batch of simulations to further scrutinize the possibility of a coincidental alignment.  Due to the slow precession rates in the extremely detached realm of the solar system, we find that an aligned system of orbits (in terms of their longitudes of perihelia) can be fossilized in some form for around 1 Gyr.  Moreover, we note that, when selecting from a finite sample of orbits (e.g.: those ETNOs and IOCOs currently near pericenter) there is always inherent bias when alignment angle is a free parameter.  Thus, a finite system of orbits is always likely to be ``clustered'' if the observer is allowed to choose the clustering angle.

In conclusion, while the Planet Nine Hypothesis \citep{vp113,batygin16} remains the most viable explanation for the observed orbital clustering in the distant solar system, still more detections of extreme objects are required to confidently infer its existence.

\section*{Acknowledgments}

We thank Sean Raymond for thought-provoking discussions and insight.  This material is based upon research supported by the Chateaubriand Fellowship of the Office for Science and Technology of the Embassy of France in the United States.  M.S.C. and N.A.K. thank the National Science Foundation for support under award AST-1615975.  The computing for this project was performed at the OU Supercomputing Center for Education and Research (OSCER) at the University of Oklahoma (OU).  Several sets of simulations were managed on the Nielsen Hall Network using the HTCondor software package: https://research.cs.wisc.edu/htcondor/.  Additional computation for the work described in this paper was supported by Carnegie Science's Scientific Computing Committee for High-Performance Computing (hpc.carnegiescience.edu).

\bibliographystyle{aasjournal}
\newcommand{\sci}{$Science$ }
\bibliography{niner.bib}

\newcommand{\noop}[1]{}
\begin{thebibliography}{}
\expandafter\ifx\csname natexlab\endcsname\relax\def\natexlab#1{#1}\fi
\providecommand{\url}[1]{\href{#1}{#1}}
\providecommand{\dodoi}[1]{doi:~\href{http://doi.org/#1}{\nolinkurl{#1}}}
\providecommand{\doeprint}[1]{\href{http://ascl.net/#1}{\nolinkurl{http://ascl.net/#1}}}
\providecommand{\doarXiv}[1]{\href{https://arxiv.org/abs/#1}{\nolinkurl{https://arxiv.org/abs/#1}}}

\bibitem[{{Alexandersen} {et~al.}(2016){Alexandersen}, {Gladman}, {Kavelaars},
  {Petit}, {Gwyn}, {Shankman}, \& {Pike}}]{alexandersen16}
{Alexandersen}, M., {Gladman}, B., {Kavelaars}, J.~J., {et~al.} 2016, \aj, 152,
  111, \dodoi{10.3847/0004-6256/152/5/111}

\bibitem[{{Bailey} {et~al.}(2016){Bailey}, {Batygin}, \& {Brown}}]{bailey16}
{Bailey}, E., {Batygin}, K., \& {Brown}, M.~E. 2016, \aj, 152, 126,
  \dodoi{10.3847/0004-6256/152/5/126}

\bibitem[{{Bailey} {et~al.}(2018){Bailey}, {Brown}, \& {Batygin}}]{bailey18}
{Bailey}, E., {Brown}, M.~E., \& {Batygin}, K. 2018, \aj, 156, 74,
  \dodoi{10.3847/1538-3881/aaccf4}

\bibitem[{{Bannister} {et~al.}(2016){Bannister}, {Kavelaars}, {Petit},
  {Gladman}, {Gwyn}, {Chen}, {Volk}, {Alexandersen}, {Benecchi}, {Delsanti},
  {Fraser}, {Granvik}, {Grundy}, {Guilbert-Lepoutre}, {Hestroffer}, {Ip},
  {Jakubik}, {Jones}, {Kaib}, {Kavelaars}, {Lacerda}, {Lawler}, {Lehner},
  {Lin}, {Lister}, {Lykawka}, {Monty}, {Marsset}, {Murray-Clay}, {Noll},
  {Parker}, {Pike}, {Rousselot}, {Rusk}, {Schwamb}, {Shankman}, {Sicardy},
  {Vernazza}, \& {Wang}}]{bannister16_ossos}
{Bannister}, M.~T., {Kavelaars}, J.~J., {Petit}, J.-M., {et~al.} 2016, \aj,
  152, 70, \dodoi{10.3847/0004-6256/152/3/70}

\bibitem[{{Bannister} {et~al.}(2017){Bannister}, {Shankman}, {Volk}, {Chen},
  {Kaib}, {Gladman}, {Jakubik}, {Kavelaars}, {Fraser}, {Schwamb}, {Petit},
  {Wang}, {Gwyn}, {Alexandersen}, \& {Pike}}]{bannister27}
{Bannister}, M.~T., {Shankman}, C., {Volk}, K., {et~al.} 2017, \aj, 153, 262,
  \dodoi{10.3847/1538-3881/aa6db5}

\bibitem[{{Bannister} {et~al.}(2018){Bannister}, {Gladman}, {Kavelaars},
  {Petit}, {Volk}, {Chen}, {Alexand ersen}, {Gwyn}, {Schwamb}, {Ashton},
  {Benecchi}, {Cabral}, {Dawson}, {Delsanti}, {Fraser}, {Granvik},
  {Greenstreet}, {Guilbert-Lepoutre}, {Ip}, {Jakubik}, {Jones}, {Kaib},
  {Lacerda}, {Van Laerhoven}, {Lawler}, {Lehner}, {Lin}, {Lykawka}, {Marsset},
  {Murray-Clay}, {Pike}, {Rousselot}, {Shankman}, {Thirouin}, {Vernazza}, \&
  {Wang}}]{bannister18}
{Bannister}, M.~T., {Gladman}, B.~J., {Kavelaars}, J.~J., {et~al.} 2018, \apjs,
  236, 18, \dodoi{10.3847/1538-4365/aab77a}

\bibitem[{{Batygin} {et~al.}(2019){Batygin}, {Adams}, {Brown}, \&
  {Becker}}]{batygin19_rev}
{Batygin}, K., {Adams}, F.~C., {Brown}, M.~E., \& {Becker}, J.~C. 2019,
  \physrep, 805, 1, \dodoi{10.1016/j.physrep.2019.01.009}

\bibitem[{{Batygin} \& {Brown}(2016{\natexlab{a}})}]{batygin16}
{Batygin}, K., \& {Brown}, M.~E. 2016{\natexlab{a}}, \aj, 151, 22,
  \dodoi{10.3847/0004-6256/151/2/22}

\bibitem[{{Batygin} \& {Brown}(2016{\natexlab{b}})}]{batygin16b}
---. 2016{\natexlab{b}}, \apjl, 833, L3, \dodoi{10.3847/2041-8205/833/1/L3}

\bibitem[{{Batygin} \& {Morbidelli}(2017)}]{batygin_morby17}
{Batygin}, K., \& {Morbidelli}, A. 2017, \aj, 154, 229,
  \dodoi{10.3847/1538-3881/aa937c}

\bibitem[{{Becker} {et~al.}(2017){Becker}, {Adams}, {Khain}, {Hamilton}, \&
  {Gerdes}}]{becker17}
{Becker}, J.~C., {Adams}, F.~C., {Khain}, T., {Hamilton}, S.~J., \& {Gerdes},
  D. 2017, \aj, 154, 61, \dodoi{10.3847/1538-3881/aa7aa2}

\bibitem[{{Becker} {et~al.}(2018){Becker}, {Khain}, {Hamilton}, {Adams},
  {Gerdes}, {Zullo}, {Franson}, {Millholland}, {Bernstein}, {Sako},
  {Bernardinelli}, {Napier}, {Markwardt}, {Lin}, {Wester}, {Abdalla}, {Allam},
  {Annis}, {Avila}, {Bertin}, {Brooks}, {Carnero Rosell}, {Carrasco Kind},
  {Carretero}, {Cunha}, {D'Andrea}, {da Costa}, {Davis}, {De Vicente}, {Diehl},
  {Doel}, {Eifler}, {Flaugher}, {Fosalba}, {Frieman}, {Garc{\'\i}a-Bellido},
  {Gaztanaga}, {Gruen}, {Gruendl}, {Gschwend}, {Gutierrez}, {Hartley},
  {Hollowood}, {Honscheid}, {James}, {Kuehn}, {Kuropatkin}, {Maia}, {March},
  {Marshall}, {Menanteau}, {Miquel}, {Ogando}, {Plazas}, {Sanchez}, {Scarpine},
  {Schindler}, {Sevilla-Noarbe}, {Smith}, {Smith}, {Soares-Santos}, {Sobreira},
  {Suchyta}, {Swanson}, {Walker}, \& {DES Collaboration}}]{becker18}
{Becker}, J.~C., {Khain}, T., {Hamilton}, S.~J., {et~al.} 2018, \aj, 156, 81,
  \dodoi{10.3847/1538-3881/aad042}

\bibitem[{{Brasser} {et~al.}(2006){Brasser}, {Duncan}, \&
  {Levison}}]{brasser06}
{Brasser}, R., {Duncan}, M.~J., \& {Levison}, H.~F. 2006, \icarus, 184, 59,
  \dodoi{10.1016/j.icarus.2006.04.010}

\bibitem[{{Brasser} {et~al.}(2012){Brasser}, {Schwamb}, {Lykawka}, \&
  {Gomes}}]{brasser12}
{Brasser}, R., {Schwamb}, M.~E., {Lykawka}, P.~S., \& {Gomes}, R.~S. 2012,
  \mnras, 420, 3396, \dodoi{10.1111/j.1365-2966.2011.20264.x}

\bibitem[{{Brown}(2001)}]{brown01}
{Brown}, M.~E. 2001, \aj, 121, 2804, \dodoi{10.1086/320391}

\bibitem[{{Brown}(2017)}]{brown17}
---. 2017, \aj, 154, 65, \dodoi{10.3847/1538-3881/aa79f4}

\bibitem[{{Brown} \& {Batygin}(2016)}]{brown16}
{Brown}, M.~E., \& {Batygin}, K. 2016, \apjl, 824, L23,
  \dodoi{10.3847/2041-8205/824/2/L23}

\bibitem[{{Brown} \& {Batygin}(2019)}]{brown19}
---. 2019, \aj, 157, 62, \dodoi{10.3847/1538-3881/aaf051}

\bibitem[{{Brown} {et~al.}(2004){Brown}, {Trujillo}, \& {Rabinowitz}}]{sedna}
{Brown}, M.~E., {Trujillo}, C., \& {Rabinowitz}, D. 2004, \apj, 617, 645,
  \dodoi{10.1086/422095}

\bibitem[{{Chambers}(1999)}]{chambers99}
{Chambers}, J.~E. 1999, \mnras, 304, 793,
  \dodoi{10.1046/j.1365-8711.1999.02379.x}

\bibitem[{{Chen} {et~al.}(2016){Chen}, {Lin}, {Holman}, {Payne}, {Fraser},
  {Lacerda}, {Ip}, {Chen}, {Kudritzki}, {Jedicke}, {Wainscoat}, {Tonry},
  {Magnier}, {Waters}, {Kaiser}, {Wang}, \& {Lehner}}]{kt19}
{Chen}, Y.-T., {Lin}, H.~W., {Holman}, M.~J., {et~al.} 2016, \apjl, 827, L24,
  \dodoi{10.3847/2041-8205/827/2/L24}

\bibitem[{{Dark Energy Survey Collaboration} {et~al.}(2016){Dark Energy Survey
  Collaboration}, {Abbott}, {Abdalla}, {Aleksi{\'c}}, {Allam}, {Amara},
  {Bacon}, {Balbinot}, {Banerji}, {Bechtol}, {Benoit-L{\'e}vy}, {Bernstein},
  {Bertin}, {Blazek}, {Bonnett}, {Bridle}, {Brooks}, {Brunner}, {Buckley-Geer},
  {Burke}, {Caminha}, {Capozzi}, {Carlsen}, {Carnero-Rosell}, {Carollo},
  {Carrasco-Kind}, {Carretero}, {Castander}, {Clerkin}, {Collett}, {Conselice},
  {Crocce}, {Cunha}, {D'Andrea}, {da Costa}, {Davis}, {Desai}, {Diehl},
  {Dietrich}, {Dodelson}, {Doel}, {Drlica-Wagner}, {Estrada}, {Etherington},
  {Evrard}, {Fabbri}, {Finley}, {Flaugher}, {Foley}, {Fosalba}, {Frieman},
  {Garc{\'\i}a-Bellido}, {Gaztanaga}, {Gerdes}, {Giannantonio}, {Goldstein},
  {Gruen}, {Gruendl}, {Guarnieri}, {Gutierrez}, {Hartley}, {Honscheid}, {Jain},
  {James}, {Jeltema}, {Jouvel}, {Kessler}, {King}, {Kirk}, {Kron}, {Kuehn},
  {Kuropatkin}, {Lahav}, {Li}, {Lima}, {Lin}, {Maia}, {Makler}, {Manera},
  {Maraston}, {Marshall}, {Martini}, {McMahon}, {Melchior}, {Merson}, {Miller},
  {Miquel}, {Mohr}, {Morice-Atkinson}, {Naidoo}, {Neilsen}, {Nichol}, {Nord},
  {Ogando}, {Ostrovski}, {Palmese}, {Papadopoulos}, {Peiris}, {Peoples},
  {Percival}, {Plazas}, {Reed}, {Refregier}, {Romer}, {Roodman}, {Ross},
  {Rozo}, {Rykoff}, {Sadeh}, {Sako}, {S{\'a}nchez}, {Sanchez}, {Santiago},
  {Scarpine}, {Schubnell}, {Sevilla-Noarbe}, {Sheldon}, {Smith}, {Smith},
  {Soares-Santos}, {Sobreira}, {Soumagnac}, {Suchyta}, {Sullivan}, {Swanson},
  {Tarle}, {Thaler}, {Thomas}, {Thomas}, {Tucker}, {Vieira}, {Vikram},
  {Walker}, {Wechsler}, {Weller}, {Wester}, {Whiteway}, {Wilcox}, {Yanny},
  {Zhang}, \& {Zuntz}}]{darkenergy16}
{Dark Energy Survey Collaboration}, {Abbott}, T., {Abdalla}, F.~B., {et~al.}
  2016, \mnras, 460, 1270, \dodoi{10.1093/mnras/stw641}

\bibitem[{{Dones} {et~al.}(2004){Dones}, {Weissman}, {Levison}, \&
  {Duncan}}]{dones04}
{Dones}, L., {Weissman}, P.~R., {Levison}, H.~F., \& {Duncan}, M.~J. 2004,
  {Oort cloud formation and dynamics}, ed. M.~C. {Festou}, H.~U. {Keller}, \&
  H.~A. {Weaver}, 153

\bibitem[{{Feng} \& {Bailer-Jones}(2015)}]{fabo15}
{Feng}, F., \& {Bailer-Jones}, C.~A.~L. 2015, \mnras, 454, 3267,
  \dodoi{10.1093/mnras/stv2222}

\bibitem[{{Fern{\'a}ndez} \& {Brunini}(2000)}]{fernandez00}
{Fern{\'a}ndez}, J.~A., \& {Brunini}, A. 2000, \icarus, 145, 580,
  \dodoi{10.1006/icar.2000.6348}

\bibitem[{{Gladman} {et~al.}(2002){Gladman}, {Holman}, {Grav}, {Kavelaars},
  {Nicholson}, {Aksnes}, \& {Petit}}]{gladman02}
{Gladman}, B., {Holman}, M., {Grav}, T., {et~al.} 2002, \icarus, 157, 269,
  \dodoi{10.1006/icar.2002.6860}

\bibitem[{{Gladman} {et~al.}(2009){Gladman}, {Kavelaars}, {Petit}, {Ashby},
  {Parker}, {Coffey}, {Jones}, {Rousselot}, \& {Mousis}}]{kv42}
{Gladman}, B., {Kavelaars}, J., {Petit}, J.-M., {et~al.} 2009, \apjl, 697, L91,
  \dodoi{10.1088/0004-637X/697/2/L91}

\bibitem[{{Gomes} {et~al.}(2017){Gomes}, {Deienno}, \& {Morbidelli}}]{gomes17}
{Gomes}, R., {Deienno}, R., \& {Morbidelli}, A. 2017, \aj, 153, 27,
  \dodoi{10.3847/1538-3881/153/1/27}

\bibitem[{{Gomes} {et~al.}(2005){Gomes}, {Levison}, {Tsiganis}, \&
  {Morbidelli}}]{gomes05}
{Gomes}, R., {Levison}, H.~F., {Tsiganis}, K., \& {Morbidelli}, A. 2005, \nat,
  435, 466, \dodoi{10.1038/nature03676}

\bibitem[{{Heisler} \& {Tremaine}(1986)}]{heisler86}
{Heisler}, J., \& {Tremaine}, S. 1986, \icarus, 65, 13,
  \dodoi{10.1016/0019-1035(86)90060-6}

\bibitem[{{Kaib} \& {Quinn}(2009)}]{kaib_quinn_sci_09}
{Kaib}, N.~A., \& {Quinn}, T. 2009, Science, 325, 1234,
  \dodoi{10.1126/science.1172676}

\bibitem[{{Kaib} {et~al.}(2011){Kaib}, {Ro{\v{s}}kar}, \& {Quinn}}]{kaib11}
{Kaib}, N.~A., {Ro{\v{s}}kar}, R., \& {Quinn}, T. 2011, \icarus, 215, 491,
  \dodoi{10.1016/j.icarus.2011.07.037}

\bibitem[{{Kaib} \& {Sheppard}(2016)}]{kaib16}
{Kaib}, N.~A., \& {Sheppard}, S.~S. 2016, \aj, 152, 133,
  \dodoi{10.3847/0004-6256/152/5/133}

\bibitem[{{Kaib} {et~al.}(2009){Kaib}, {Becker}, {Jones}, {Puckett}, {Bizyaev},
  {Dilday}, {Frieman}, {Oravetz}, {Pan}, {Quinn}, {Schneider}, \&
  {Watters}}]{kaib09}
{Kaib}, N.~A., {Becker}, A.~C., {Jones}, R.~L., {et~al.} 2009, \apj, 695, 268,
  \dodoi{10.1088/0004-637X/695/1/268}

\bibitem[{{Kaib} {et~al.}(2019){Kaib}, {Pike}, {Lawler}, {Kovalik}, {Brown},
  {Alexandersen}, {Bannister}, {Gladman}, \& {Petit}}]{kaib19}
{Kaib}, N.~A., {Pike}, R., {Lawler}, S., {et~al.} 2019, \aj, 158, 43,
  \dodoi{10.3847/1538-3881/ab2383}

\bibitem[{{Kavelaars} {et~al.}(2020){Kavelaars}, {Lawler}, {Bannister}, \&
  {Shankman}}]{kavelaars19}
{Kavelaars}, J.~J., {Lawler}, S.~M., {Bannister}, M.~T., \& {Shankman}, C.
  2020, {Perspectives on the distribution of orbits of distant Trans-Neptunian
  objects}, ed. D.~{Prialnik}, M.~A. {Barucci}, \& L.~{Young}, 61--77,
  \dodoi{10.1016/B978-0-12-816490-7.00003-5}

\bibitem[{{Khain} {et~al.}(2018){Khain}, {Batygin}, \& {Brown}}]{khain18}
{Khain}, T., {Batygin}, K., \& {Brown}, M.~E. 2018, \aj, 155, 250,
  \dodoi{10.3847/1538-3881/aac212}

\bibitem[{{Levison} {et~al.}(2001){Levison}, {Dones}, \& {Duncan}}]{levison01}
{Levison}, H.~F., {Dones}, L., \& {Duncan}, M.~J. 2001, \aj, 121, 2253,
  \dodoi{10.1086/319943}

\bibitem[{{Levison} \& {Duncan}(1994)}]{levison94}
{Levison}, H.~F., \& {Duncan}, M.~J. 1994, \icarus, 108, 18,
  \dodoi{10.1006/icar.1994.1039}

\bibitem[{{Levison} {et~al.}(2008){Levison}, {Morbidelli}, {Van Laerhoven},
  {Gomes}, \& {Tsiganis}}]{levison08}
{Levison}, H.~F., {Morbidelli}, A., {Van Laerhoven}, C., {Gomes}, R., \&
  {Tsiganis}, K. 2008, \icarus, 196, 258, \dodoi{10.1016/j.icarus.2007.11.035}

\bibitem[{{Li} {et~al.}(2018){Li}, {Hadden}, {Payne}, \& {Holman}}]{li18}
{Li}, G., {Hadden}, S., {Payne}, M., \& {Holman}, M.~J. 2018, \aj, 156, 263,
  \dodoi{10.3847/1538-3881/aae83b}

\bibitem[{{Malhotra} {et~al.}(2016){Malhotra}, {Volk}, \& {Wang}}]{malhotra16}
{Malhotra}, R., {Volk}, K., \& {Wang}, X. 2016, \apjl, 824, L22,
  \dodoi{10.3847/2041-8205/824/2/L22}

\bibitem[{{Millholland} \& {Laughlin}(2017)}]{millholland17}
{Millholland}, S., \& {Laughlin}, G. 2017, \aj, 153, 91,
  \dodoi{10.3847/1538-3881/153/3/91}

\bibitem[{{Morbidelli}(2002)}]{morby02_book}
{Morbidelli}, A. 2002, {Modern celestial mechanics : aspects of solar system
  dynamics}

\bibitem[{{Morbidelli} {et~al.}(2008){Morbidelli}, {Levison}, \&
  {Gomes}}]{morby08}
{Morbidelli}, A., {Levison}, H.~F., \& {Gomes}, R. 2008, {The Dynamical
  Structure of the Kuiper Belt and Its Primordial Origin}, ed. M.~A. {Barucci},
  H.~{Boehnhardt}, D.~P. {Cruikshank}, A.~{Morbidelli}, \& R.~{Dotson}, 275

\bibitem[{{Morbidelli} {et~al.}(2005){Morbidelli}, {Levison}, {Tsiganis}, \&
  {Gomes}}]{mor05}
{Morbidelli}, A., {Levison}, H.~F., {Tsiganis}, K., \& {Gomes}, R. 2005, \nat,
  435, 462, \dodoi{10.1038/nature03540}

\bibitem[{{Morbidelli} \& {Nesvorn{\'y}}(2020)}]{morb_nes_rev19}
{Morbidelli}, A., \& {Nesvorn{\'y}}, D. 2020, {Kuiper belt: formation and
  evolution}, ed. D.~{Prialnik}, M.~A. {Barucci}, \& L.~{Young}, 25--59,
  \dodoi{10.1016/B978-0-12-816490-7.00002-3}

\bibitem[{{Murray} \& {Dermott}(1999)}]{murray_dermott}
{Murray}, C.~D., \& {Dermott}, S.~F. 1999, {Solar system dynamics}

\bibitem[{{Nesvorn{\'y}} {et~al.}(2017){Nesvorn{\'y}}, {Vokrouhlick{\'y}},
  {Dones}, {Levison}, {Kaib}, \& {Morbidelli}}]{nesvorny17}
{Nesvorn{\'y}}, D., {Vokrouhlick{\'y}}, D., {Dones}, L., {et~al.} 2017, \apj,
  845, 27, \dodoi{10.3847/1538-4357/aa7cf6}

\bibitem[{{Oort}(1927)}]{oort27}
{Oort}, J.~H. 1927, \bain, 3, 275

\bibitem[{{Petit} {et~al.}(2011){Petit}, {Kavelaars}, {Gladman}, {Jones},
  {Parker}, {Van Laerhoven}, {Nicholson}, {Mars}, {Rousselot}, {Mousis},
  {Marsden}, {Bieryla}, {Taylor}, {Ashby}, {Benavidez}, {Campo Bagatin}, \&
  {Bernabeu}}]{petit11}
{Petit}, J.~M., {Kavelaars}, J.~J., {Gladman}, B.~J., {et~al.} 2011, \aj, 142,
  131, \dodoi{10.1088/0004-6256/142/4/131}

\bibitem[{{Petit} {et~al.}(2017){Petit}, {Kavelaars}, {Gladman}, {Jones},
  {Parker}, {Bieryla}, {Van Laerhoven}, {Pike}, {Nicholson}, {Ashby}, \&
  {Lawler}}]{petit17}
---. 2017, \aj, 153, 236, \dodoi{10.3847/1538-3881/aa6aa5}

\bibitem[{{Reid} {et~al.}(2002){Reid}, {Gizis}, \& {Hawley}}]{reid02_PDMF}
{Reid}, I.~N., {Gizis}, J.~E., \& {Hawley}, S.~L. 2002, \aj, 124, 2721,
  \dodoi{10.1086/343777}

\bibitem[{{Rickman} {et~al.}(2008){Rickman}, {Fouchard}, {Froeschl{\'e}}, \&
  {Valsecchi}}]{rickman08}
{Rickman}, H., {Fouchard}, M., {Froeschl{\'e}}, C., \& {Valsecchi}, G.~B. 2008,
  Celestial Mechanics and Dynamical Astronomy, 102, 111,
  \dodoi{10.1007/s10569-008-9140-y}

\bibitem[{{Saillenfest}(2020{\natexlab{a}})}]{saillenfest20_review}
{Saillenfest}, M. 2020{\natexlab{a}}, arXiv e-prints, arXiv:2001.07579.
\newblock \doarXiv{2001.07579}

\bibitem[{{Saillenfest}(2020{\natexlab{b}})}]{saillenfest19}
---. 2020{\natexlab{b}}, Celestial Mechanics and Dynamical Astronomy, 132, 12,
  \dodoi{10.1007/s10569-020-9954-9}

\bibitem[{{Shankman} {et~al.}(2017){Shankman}, {Kavelaars}, {Bannister},
  {Gladman}, {Lawler}, {Chen}, {Jakubik}, {Kaib}, {Alexandersen}, {Gwyn},
  {Petit}, \& {Volk}}]{shankman17}
{Shankman}, C., {Kavelaars}, J.~J., {Bannister}, M.~T., {et~al.} 2017, \aj,
  154, 50, \dodoi{10.3847/1538-3881/aa7aed}

\bibitem[{{Sheppard} {et~al.}(2019){Sheppard}, {Trujillo}, {Tholen}, \&
  {Kaib}}]{sheppard19}
{Sheppard}, S.~S., {Trujillo}, C.~A., {Tholen}, D.~J., \& {Kaib}, N. 2019, \aj,
  157, 139, \dodoi{10.3847/1538-3881/ab0895}

\bibitem[{{Trujillo} \& {Sheppard}(2014)}]{vp113}
{Trujillo}, C.~A., \& {Sheppard}, S.~S. 2014, \nat, 507, 471,
  \dodoi{10.1038/nature13156}

\bibitem[{{Tsiganis} {et~al.}(2005){Tsiganis}, {Gomes}, {Morbidelli}, \&
  {Levison}}]{Tsi05}
{Tsiganis}, K., {Gomes}, R., {Morbidelli}, A., \& {Levison}, H.~F. 2005, \nat,
  435, 459, \dodoi{10.1038/nature03539}

\end{thebibliography}
\end{document}